\documentclass[letterpaper,12pt]{article}

\usepackage{jheppub}

\usepackage{amsmath}
\usepackage{graphicx}
\usepackage{mathtools}
\usepackage{subfig}

\newcommand{\dlog}{d\mathrm{log}}
\newcommand{\Res}{\mathrm{Res}}

\newcommand{\pathfig}{Figures}
\newcommand{\addpic}[1]{\left(\vcenter{\hbox{\includegraphics[width=2cm]{Figures/#1}}} \right)}
\newcommand{\addpicwide}[1]{\left(\vcenter{\hbox{\includegraphics[width=3cm,trim={.55cm 0cm .5cm 0cm}]{Figures/#1}}} \right)}
\newcommand{\eref}[1]{(\ref{#1})}

\title{Scattering Amplitudes and Simple Canonical Forms for Simple Polytopes}
\author{Giulio Salvatori}
\author{and Stefan Stanojevic}

\emailAdd{giulio\_salvatori@brown.edu}
\emailAdd{stefan\_stanojevic@alumni.brown.edu}

\affiliation{Department of Physics, Brown University, Providence RI 02912, USA}

\abstract{We provide an efficient recursive formula to compute the canonical forms of arbitrary $d$-dimensional \emph{simple} polytopes, which are convex polytopes such that  every vertex lies precisely on $d$ facets. For illustration purposes, we explicitly derive recursive formulae for the canonical forms of Stokes polytopes, which play a similar role for a theory with quartic interaction as the Associahedron does in planar bi-adjoint $\phi^3$ theory.
As a by-product, our formula also suggests a new way to obtain the full planar amplitude in $\phi^4$ theory by taking suitable limits of the canonical forms of constituent Stokes polytopes.}

\begin{document}
  \maketitle

\section{Introduction}

Scattering amplitudes are among the most fundamental objects in physics and therefore it is perhaps not surprising that their study often reveals connections with profound ideas in disparate branches of mathematics.
A paradigmatic example of these connections is the discovery that tree-level amplitudes and loop-integrands in planar $\mathcal{N}=4$ SYM are intimately tied to a geometrical object called the \emph{Amplituhedron} \cite{Arkani-Hamed:2013jha, ArkaniHamed:2012nw}.
Roughly speaking, the tree-level Amplituhedron can be thought of as a Grassmannian generalization of the convex hull of external kinematical data, and amplitudes are extracted from the unique differential form with simple poles at its boundaries. In particular, the two very distinctive features of the analytic structure of amplitudes, often referred to as \emph{Locality} and \emph{Unitarity}, emerge from the boundary structure of the Amplituhedron which, in turn, is implied in a very non-trivial way by the ``convex hull'' construction defining it.
In this picture the computation of amplitudes is thus translated into the geometrical problem of characterizing the boundary and the interior of the Amplituhedron, a problem which, while being unfamiliar in physics, is closely connected to similar questions in combinatorics, algebraic geometry and cluster algebras.

In the last few years it has been understood that this novel picture for scattering amplitudes extends beyond the very special $\mathcal{N}=4$ SYM theory. Amplituhedra have been found to underlie amplitudes in bi-adjoint scalar theory at tree level \cite{Arkani-Hamed:2017mur}, integrands at 1-loop level \cite{Salvatori:2018aha}, and constituents of planar tree level amplitudes in a host of scalar theories with polynomial interactions \cite{Banerjee:2018tun},\cite{Raman:2019utu},\cite{Jagadale:2019byr}, as well as in non-planar ones \cite{Gao:2017dek}.
It is promising to discover that the same set of ideas can be used to describe amplitudes in theories as vastly different as $\mathcal{N}=4$ SYM and bi-adjoint $\phi^3$.
On the other hand, it is perhaps not surprising that the role of the Amplituhedron in these elementary theories is played directly by convex polytopes which, in the language of \cite{Arkani-Hamed:2017tmz}, are the most basic instances of \emph{Positive Geometries}, whereas the $\mathcal{N}=4$ SYM Amplituhedron sits on the wildest end of the spectrum of these geometries.

It is probably wise at this point to caution the reader, especially the physicist reader, that the elementary nature of convex polytopes can be deceptive. The study of polytopes is indeed a rich branch of mathematics, with connections to combinatorics, geometry and abstract algebra, where foundational results were established only recently. Just to give an example, to this day it is not understood whether a putative ``combinatorial" polytope can be realized as the face lattice of an actual convex polytope.
Because of this, the existence of infinite families of polytopes associated with the theories described above should be appreciated for being a very non-trivial fact.

A common trait of the newly discovered Amplituhedra is that they all are \emph{simple} polytopes, which means that each of their vertices lies at the intersection of precisely $d$ facets \footnote{By facet we mean a codimension one boundary of the polytope}, $d$ being the dimension of the polytope itself.
The canonical form of these polytopes can be immediately written, following \cite{Arkani-Hamed:2017mur}, as a sum over its vertices
\begin{align}%\eqref{eq:canonicalsimple}
    \Omega = \sum_{v \in \mathrm{vertices}} \mathrm{sgn}(v) \bigwedge_{\substack{f \in \mathrm{facets}  \\ v \in f }} d\mathrm{log}(X_f),
    \label{eq:canonicalsimple}
\end{align}
where $X_f$ is the variable associated to a facet $f$. When a convex realization of the polytope is known, all the variables $X_f$ are given by affine linear functions. However, it will be enlightening for the moment to think of the $X_f$ as independent variables and $\Omega$ as a differential form defined on the space spanned by all of them, which we collectively call $X$.
We still have to decide how to assign the relative signs of the $d\mathrm{log}$s appearing in \eqref{eq:canonicalsimple}. This can be done uniquely up to an overall sign by requiring $\Omega$ to be \emph{projective}, i.e. invariant under local rescaling of the variables $X \to \alpha(X) X$ for any rational function $\alpha(X)$. We can give an appealing interpretation for the projectivity of a form by introducing an operator $\delta$ which captures the variation of a form under such rescaling. The operator $\delta$ is defined on any differential form $\Omega$ which can be written as a linear combination of $d\mathrm{log}$s by noting that its variation under $X \to \alpha(X) X$ is proportional to $d\mathrm{log}(\alpha)$, we define $\delta(\Omega)$ to be the form obtained by stripping this factor:
$$\dlog{\alpha}\wedge \delta(\Omega) := \left.\Omega\right|_{X \to \alpha(X)} - \Omega.$$
The projective variation operator $\delta$ satisfies $\delta^2 = 0$. This can be proven by first showing that $\delta^2(\Omega) = 0$ if $\Omega$ consists of a single $d\mathrm{log}$ term. By the linearity of $\delta$, the result is then extended to any differential form $\Omega$ written as a linear combination of $d\mathrm{log}$ terms. On a single $d\mathrm{log}$ term $\delta$ acts as
$$\delta( \dlog(X_0) \wedge \dots \wedge \dlog(X_n)) = \ \sum_{i=0}^n (-1)^i \dlog(X_0) \wedge \dots  \wedge \widehat{\dlog(X_i)}\wedge \dots \wedge \dlog(X_n), $$
where the hat denotes a missing factor. By further applying $\delta$ it easily follows that $\delta^2(\Omega) = 0$.
Let $\Omega$ be any differential form, not necessarily projective, written as a combination of $d\mathrm{log}$s. Consider the form
\begin{align}
  \tilde{\Omega} = \Omega - d\mathrm{log}(M) \wedge \delta(\Omega),
  \label{eq:tildeOmega}
\end{align}
where $M$ is a new variable not already appearing in $\Omega$. From the projectivity of $\delta(\Omega)$ it follows that $\tilde{\Omega}$ is projective. Indeed, computing the trasformation of the RHS of \eqref{eq:tildeOmega} one gets
$$\tilde{\Omega} = \Omega - d\mathrm{log}(M) \wedge \delta(\Omega) \to \Omega + d\mathrm{log}(\alpha) \wedge \delta(\Omega) - (\dlog(\alpha)+\dlog(M))\wedge\delta(\Omega) = \tilde{\Omega}$$
It follows that $\tilde{\Omega}$ is well defined as a top form on the projective space with homogeneous coordinates $Y = (M,X)$. Clearly, $\tilde{\Omega}$ reduces to $\Omega$ in the affine chart where $M=1$. However, $\tilde{\Omega}$ also develops a pole along the plane at infinity $M=0$ with residue $\delta(\Omega)$. In conclusion, the projectivity of $\Omega$ is equivalent to the absence of a pole at infinity when the variables $X$ are projectivized.

For a simple polytope the projectivity of $\Omega$ is a remarkable property which is, however, obscured in \eqref{eq:canonicalsimple}: the $d\mathrm{log}$ terms are not individually invariant and the projectivity is hidden in the fact that relative signs can be coherently assigned to all vertices in such a way that their projective variations cancel each other. En passant we recall that vertices are associated to Feynman diagrams, so we can rephrase this fact as saying that there is a symmetry of the amplitude obscured term-by-term in the Feynman diagramatic expansion \footnote{Building on this in \cite{Arkani-Hamed:2017mur} an intriguing analogy between projectivity of the canonical form and dual conformal invariance in $\mathcal{N}=4$ SYM was drawn.}.
It is natural to try and make manifest this symmetry by finding a representation for $\Omega$ which is manifestly invariant under $\delta$.

We will see shortly that such a representation exists, but first let us start with a few illustrative examples.
The simplest polytope is a segment for which the canonical form can be trivially made manifestly projective,
\begin{align}
    \Omega_2 = \dlog(X_1) - \dlog(X_2) = \dlog\left(\frac{X_1}{X_2}\right),
    \label{eq:segment}
\end{align}
since it depends only on the ratio $X_1/X_2$.

\noindent
\begin{figure}
    \centering
    \minipage{0.5\textwidth}
    \centering
    \includegraphics[trim={5cm 3.2cm 4.4cm 4cm},width=.85\textwidth]{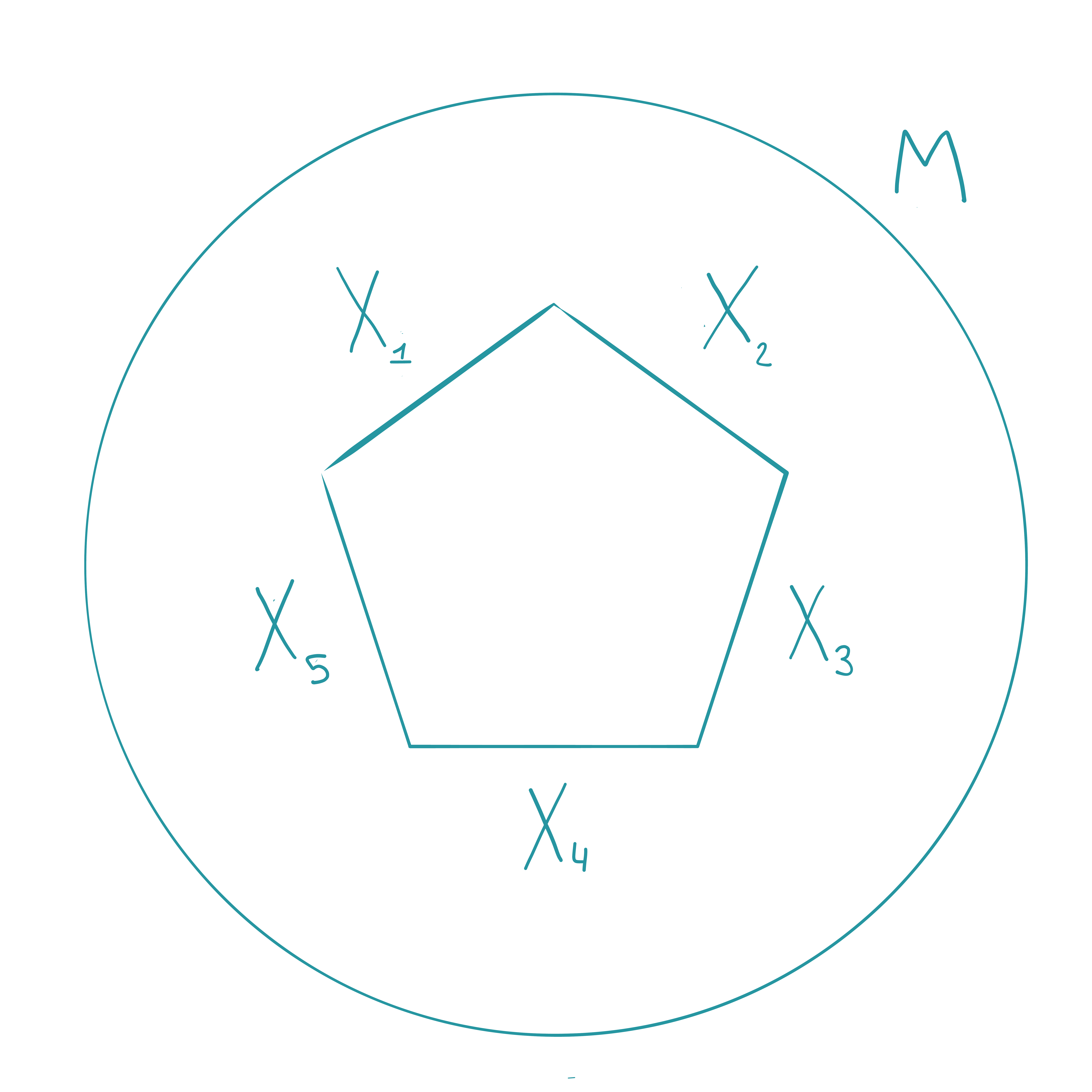}
    \endminipage\hfill
    \minipage{0.5\textwidth}
    \centering
    \includegraphics[trim={30cm 28cm 28cm 33cm},width=.85\textwidth]{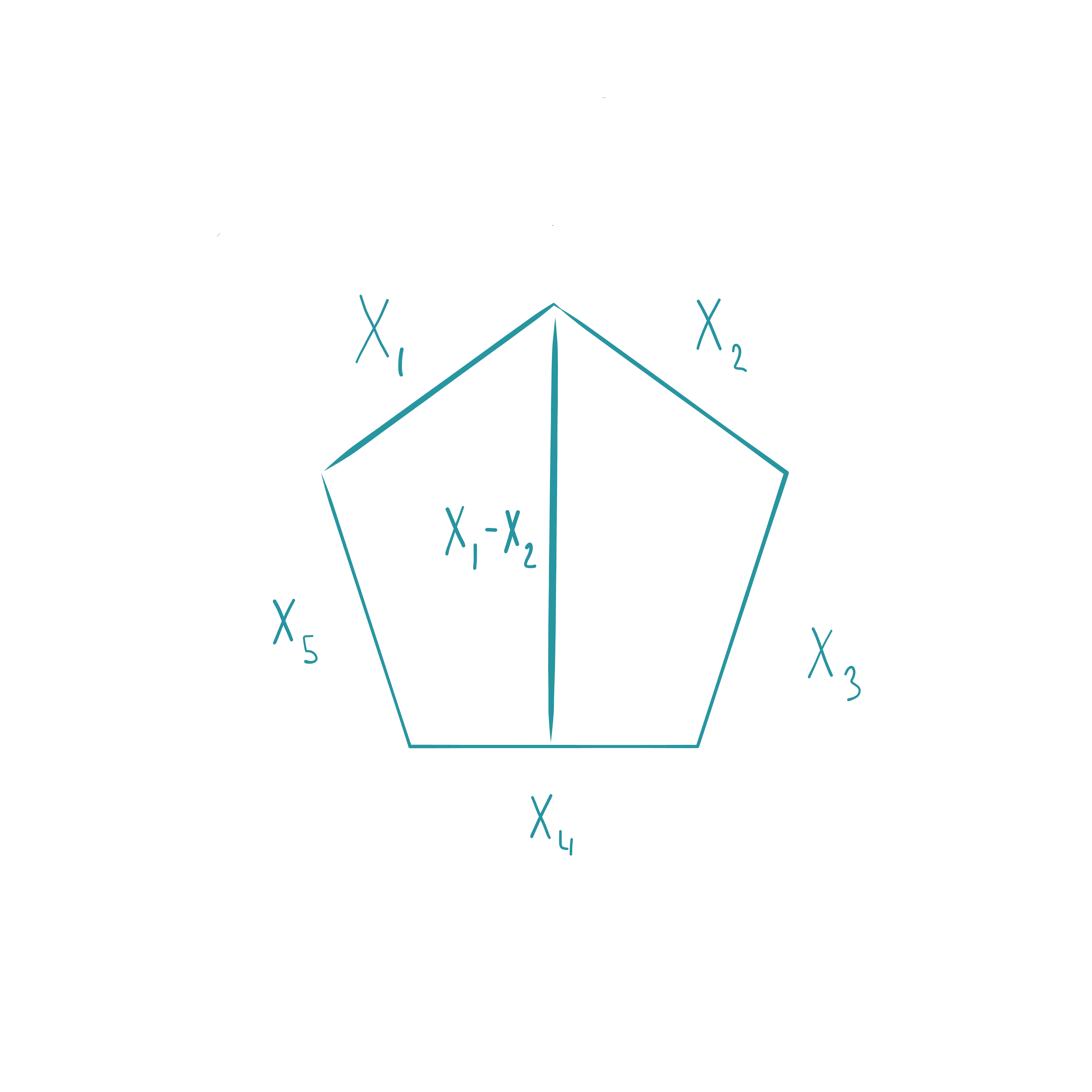}
    \endminipage\hfill
    \caption{A pentagon with the line at infinity $\{M = 0\}$ (left) and its decomposition into two squares (right).}
    \label{fig:pentagon}
\end{figure}
\noindent
The first non-trivial shape that we encounter is a 2-dimensional pentagon as in Fig. \ref{fig:pentagon} whose canonical form is
\begin{align}
    \Omega_5 &= \dlog(X_1)\dlog(X_2) - \dlog(X_3)\dlog(X_2) \nonumber \\
    &+ \dlog(X_3)\dlog(X_4) - \dlog(X_5)\dlog(X_4)+\dlog(X_5)\dlog(X_1),
\end{align}
and now there is no obvious way of recombining the $\dlog$s in a manifestly projective way. However, we can decompose the pentagon in the two squares as in Fig. \ref{fig:pentagon} and, accordingly, write its canonical form as the sum of the two forms associated to each square. We introduce an extra boundary which, in order to be internal to the pentagon, must be of the form $X_{\mathrm{spurious}} = X_1 - \epsilon X_2$ for some positive $\epsilon$; it will be sufficient to make the simplest choice $X_\mathrm{spurious} = X_1 - X_2$.
We obtain the expression
\begin{align}
    \Omega_5 &= \left[\dlog(X_1) - \dlog(X_4)\right]\wedge\left[\dlog(X_5) - \dlog(X_2-X_1)\right]+\nonumber\\
    &+\left[\dlog(X_2) - \dlog(X_4)\right]\wedge\left[\dlog(X_3) - \dlog(X_1-X_2)\right] =\nonumber\\
    &=\dlog\left(\frac{X_1}{X_4}\right)\wedge\dlog\left(\frac{X_5}{X_2-X_1}\right)+ \dlog\left(\frac{X_2}{X_4}\right)\wedge\dlog\left(\frac{X_3}{X_1-X_2}\right),
    \label{eq:triangulation}
\end{align}
which is manifestly projective.
Before going further, we wish to emphasize that in the above computation the details of the particular convex realization of the pentagon do not appear. In particular, there was no real motivation to draw the line $X_{\mathrm{spurious}}$ in Fig. \ref{fig:pentagon} as intersecting the facet $X_4$ inside of the pentagon. Nevertheless, the equality in \eqref{eq:triangulation} stands as an equality between differential forms on the space of \emph{independent} variables $X$.
We should also emphasize a fact which might have gone unnoticed in the derivation of \eqref{eq:triangulation} because of the particularly simple geometry in consideration. Each of the squares can be thought of as a $prism$, by which we mean a polytope with two combinatorically equivalent facets -  let us call them $\mathcal{U}$p and $\mathcal{B}$ottom -  plus many $\mathcal{S}$ide facets joining them. Because of its simple structure, it is easy to guess that the canonical form of a prism is
\begin{align}
\Omega_{\mathcal{P}} = \dlog\left(\frac{X_\mathcal{B}}{X_\mathcal{U}} \right) \wedge D_u \Omega_\mathcal{U},
\label{eq:naiveprism}
\end{align}
where $\Omega_\mathcal{U}$ is the canonical form of the upper facet (we could also choose to use the bottom facet) and $D_u$ is an operator, whose precise form will be given later, that must promote the poles that in $\Omega_{\mathcal{U}}$ are associated to the intersections $\mathcal{U} \cap \mathcal{S}$ to poles along the higher dimensional facets $\mathcal{S}$.
Therefore in \eqref{eq:triangulation} we are both recursively computing $\Omega_5$, by recycling the result of \eqref{eq:segment}, and making it manifestly projective in one fell swoop.

This is not the end of the story. The newly discovered Amplituhedra for $\phi^3$ \footnote{For $\phi^4$ theory, some examples of convex realizations of the corresponding polytopes were proposed in \cite{Banerjee:2018tun}, \cite{Aneesh:2019cvt}} theory admit a convex realization, i.e. a subspace $H$ where all the variables $X$ are given by affine linear functions, such that the pullback of $\Omega$ on this space is given by
\begin{align}%\eqref{eq:rationalsimple}
    \Omega|_{H} = dX_1 \wedge \dots \wedge dX_d \left( \sum_{v \in \mathrm{vertices}} \prod_{\substack{f \in \mathrm{facets}  \\ v \in f }} \frac{1}{X_f} \right),  
    \label{eq:rationalsimple}
\end{align}
where $(X_1, \dots, X_d)$ is an arbitrary choice of coordinates for $H$. The rational function $\underline{\Omega}$ obtained by stripping the differential form $dX_1 \wedge \dots \wedge dX_d$ from \eqref{eq:rationalsimple} is then immediately recognised as the Feynman diagramatic expansion of the amplitude.
For example, in the case of the pentagon one obtains
\begin{align}
    \underline{\Omega}_5 = \sum_{i=1}^5 \frac{1}{X_i X_{i+1} }.
    \label{eq:pentagonRational}
\end{align}
Note that the relative signs required for the projectivity of $\Omega$ get miracolously balanced by the pullback on $H$. Staring at \eqref{eq:triangulation} is then tempting to guess a recursive expression for the rational function of the pentagon
\begin{align}
    \underline{\Omega}_5 \overset{?}{=}
    \left(\frac{1}{X_1}+\frac{1}{X_4}\right)\left(\frac{1}{X_5}+\frac{1}{X_1 - X_2}\right)+\left(\frac{1}{X_2}+\frac{1}{X_4}\right)\left(\frac{1}{X_3}+\frac{1}{X_2 - X_1}\right),
    \label{eq:putative}
\end{align}
by direct comparison with \eqref{eq:pentagonRational} it is easy to see that, indeed, \eqref{eq:putative} yields an identity between rational functions in the variables $X_i$.
Once again, we wish to stress that the details of the convex realization of the pentagon drop out, and one is left with recursive formulae for amplitudes which are correct on the space of independent variables $X_i$.

%The above constructions were inspired by a remarkable projection property satisfied by the so called generalized ABHY Associahedra, which will be discussed elsewhere \cite{GiulioNima}. However, it was later understood that they hold in greater generality for arbitrary simple polytopes.
%Let us remark here that the example of the pentagon above is a special case of certain recursive formulae for the canonical form of the so called ABHY Associahedra, which were inspired by a remarkable projection property satisfied by these polytopes \cite{GiulioNima, GiulioPhD}.

In the rest of the paper we will state and prove the recursive formulae for $\Omega$ and $\underline{\Omega}$ in full generality. As an illustrative example we will apply it to Stokes polytopes obtaining novel representations for their rational functions. As we will argue later, the simple structure of the recursions also suggests a natural way to get rid of double countings of Feynman diagrams across different polytopes by taking suitable limits, a fact that we will use to obtain new expressions for the full planar amplitude as well.

This paper is structured as follows. In Section \ref{Section:review} we offer a self-contained review of the definition of positive geometries and canonical forms focused on the case of simple polytopes. In Section \ref{Section:formulae} we present and prove our main results, as well as providing a few simple examples.
In Section \ref{Section:phi4} we apply our formulae to the case of Stokes polytopes, obtaining recursive representations for their rational functions, as well as new expressions for the full $\phi^4$ amplitude. Finally, in Section \ref{sec:conclusions} we draw our conclusions.

\section{A brief review of positive geometries}
\label{Section:review}

We review in more details some of the ideas put forward in \cite{Arkani-Hamed:2017tmz} which we touched in the introduction. We then state and prove a simple recursive formula to compute the canonical form of simple polytopes. Remarkably, a naive guess allows to extend this to a recursive formula for rational functions canonically associated to simple polytopes. As we will see, however, this latter formula requires some extra combinatorical requirement on the polytope.

\subsection{Polytopes and their canonical forms}
\label{sec:polytopes}

Let $\mathcal{P}$ be a d-dimensional simple polytope. If we label the facets of $\mathcal{P}$ using variables $X_f$, $f=1, \dots, F$, then each vertex $v$ of $\mathcal{P}$ is uniquely identified by a $d$-tuple of variables $X$. Note that $F \ge d+1$, the inequality being saturated if and only if $\mathcal{P}$ is a simplex.
Without loss of generality, we assume that $\mathcal{P}$ is given a convex realization as the intersection of the positive region $\mathbb{R}^F_{\ge 0} \coloneqq \{ X_f \ge 0\}$ with an appropriate d-dimensional affine subspace $H$. Thinking projectively, we introduce a homogeneous vector $\mathcal{Y} = (M, X_1, \dots, X_F)$, then $H$ is defined by the constraint $\mathcal{C} \cdot \mathcal{Y}=0$ where $\mathcal{C}$ is a $(F-d)\times (F+1)$ matrix. The subspace $H$ can alternatively be encoded by writing each of the $F$ variables $X_f$ as an affine function, i.e. by writing $X = W \cdot Y$, where $Y$ is a $(d+1)$-vector of homogeneous coordinates for $H$. By performing a $\mathrm{GL}(d+1)$ transformation we can always choose $d$ compatible variables $X_f$ to be the affine coordinates for $H$, which means that we center the origin of our space at one of the vertices $v$ of $\mathcal{P}$. Then we have to specify only $F-d$ rows of $W$; they are the in-ward normal vectors of $F-d$ corresponding facets of $\mathcal{P}$.

To any convex polytope $\mathcal{P}$, or more generally to any positive geometry, we can uniquely associate a meromorphic top-dimensional differential form $\Omega_{\mathcal{P}}$, which is defined in a iterative way by the requirements
\begin{align}%\eqref{eq:canonicalDef}
    \Res_{X_f}(\Omega_\mathcal{P}) = \Omega_{X_f},\ \mathrm{and}\ \Omega_{\mathrm{point}} = \pm 1,
    \label{eq:canonicalDef}
\end{align}
By further requiring that $\Omega_{\mathcal{P}}$ is holomorphic elsewhere, $\Omega_P$ is fixed up to a sign.
We recall that the residue\footnote{In the mathematical literature there are several notions of multivariate generalizations of the familiar residue from complex analysis. The one invoked here is known as \emph{Poincare Residue}.}  of a top degree differential form along an affine subspace $\{X_f = 0\}$ is defined by 
\begin{align}
    \Res_{X_f}(\dlog(X_f) \wedge \omega + \eta) = \left.\omega\right|_{X_f =0}(\Omega),
    \label{eq:residue}
\end{align}
where $\omega$ (which may be zero) and $\eta$ are regular along $\{X_f = 0\}$. The residue operator yields a differential form on the subspace $\{X_f=0\}$.
The operation of taking residues can be iterated and one can define the operator \begin{align}
\Res_{(X_{f_1}, \dots, X_{f_m} )}(\Omega) \coloneqq \Res_{X_{f_1}} \dots \Res_{X_{f_m}},
\end{align}
where we stress that the functions $X_{f_i}$ have to be restricted to the subspaces on which residues have already been taken. 

As we anticipated in the introduction, the canonical form of a simple polytope ${\mathcal{P}}$ is given by
\begin{align}%\eqref{eq:canonicalsimple2}
    \Omega_{\mathcal{P}} = \sum_{v \in \mathrm{vertices}} \mathrm{sgn}(v) \bigwedge_{\substack{f \in \mathrm{facets}  \\ v \in f }} d\mathrm{log}(X_f),
    \label{eq:canonicalsimple2}
\end{align}
where the signs are fixed by projectivity. More precisely, the statement is that the pullback of $\Omega_{\mathcal{P}}$ on the subspace $H$ defining the convex realization of $\mathcal{P}$ yields its canonical form. The proof is by comparison with a known representation of $\underline{\Omega}$ as integral over the dual polytope $\mathcal{P}^*$, in this language the plane at infinity $\{M = 0\}$ is dual to a point in the interior of $\mathcal{P}^*$, which proves that neither $\Omega$ or its pullback on $H$ develops a pole there. In practice, the signs can be fixed as follows. 
Any 1-dimensional face of $\mathcal{P}$ is uniquely associated to a collection of $d-1$ facets $(X_{a_1}, \dots, X_{a_{d-1}})$ and it touches $\mathcal{P}$ at two vertices specified by two additional variables $X$ and $X'$. In order not to develop a pole at infinity along the line $(X_{a_1}, \dots, X_{a_{d-1}})$ we must have
\begin{align}
    \Res_{(X_{a_1}, \dots, X_{a_m}, X)}(\Omega) = - \Res_{(X_{a_1}, \dots, X_{a_m}, X')}(\Omega).
    \label{eq:adjacencyrule}
\end{align}
We can start from any vertex $v_0$, choose an arbitrary sign for the corresponding $\dlog$, then explore all the vertices of $\mathcal{P}$ by moving along its 1-dimensional faces and fix the signs according to \eqref{eq:adjacencyrule}. 

A priori it is not obvious that the procedure just described is going to be consistent, i.e. that the the assignment of signs will not depend on a particular path chosen to get to a far away vertex.  However, we know that this must be the case because $\mathcal{P}$ is a polytope, in particular it admits a dual $\mathcal{P}^*$ which gives an independent proof of the projectivity of $\Omega_{\mathcal{P}}$. Note, however, that the procedure only requires the knowledge of the graph of $\mathcal{P}$, i.e. the collection of its one dimensional faces.
For example, the particular subspace $H$ does not effectively appear in $\Omega$.
It is tempting to try and define a form for an arbitrary graph $G$ and it is then an interesting question to understand what are the topological properties of $G$ that guarantee the projectivity of $\Omega$.

Because the particular realization of the polytope $\mathcal{P}$ does not effectively appear in the definition of $\Omega_{\mathcal{P}}$ it is natural to study it as a differential form on the affine space generated by all the facet variables, thought of as independent variables.
This poses an obvious problem, in that the usual residue operator is well defined on top-degree forms, e.g. the residue $\Res_{y} \frac{dx}{y}$ would not make sense. However, if we restrict ourselves to forms which are given by linear combinations of $\dlog$s one can still define a well behaved $\Res$ operator. We will review this construction in the following section.

If on one hand differential forms are very natural objects to consider in order to speak of residues, on the other scattering amplitudes are ultimately functions of the kinematical data.
When dealing with top degree differential forms on a projective space the distinction is irrelevant since any such form can be written as
$\Omega = \langle Y d^d Y \rangle \underline{\Omega}(Y),$
where $\underline{\Omega}(Y)$ is an homogeneous function of weight $-(d+1)$ and $\langle Y d^d Y \rangle \coloneqq \det(Y dY \dots dY)$ is a standard measure on $\mathbb{P}^d$. Therefore we can unambiguosly pass from the differential form $\Omega$ to the rational function $\underline{\Omega}$.
However, the number of facets of any polytope $\mathcal{P}$ is always greater than its dimension so that the canonical form $\Omega_{\mathcal{P}}$ is never a top degree form. The obvious solution is to consider its pullback $\Omega_{H}$ on the space $H$ on which $\mathcal{P}$ is geometrically realized, which is now a top degree form and is thus associated to a rational function. In the case of a simple polytope from $\eqref{eq:canonicalsimple2}$ we get
\begin{align}%\eqref{eq:rationalsimple}
    \underline{\Omega}|_{H} = \left( \sum_{v \in \mathrm{vertices}} \mathrm{sgn}(v) \langle W_M W_{f_1} \dots W_{f_d} \rangle  \prod_{\substack{f \in \mathrm{facets}  \\ v \in f }} \frac{1}{X_f} \right),  
    \label{eq:withnumerators}
\end{align}
where $(X_1, \dots, X_d)$ are affine coordinates for $H$, the variables $X_f$ are now given by linear functions $X_f = W_f \cdot (1, X_1, \dots, X_d)$ and $W_M = (1, 0, \dots, 0)$.
Furthermore, in all cases encountered so far, the pullback space $H$ is such that the signs $\mathrm{sgn}(v)$ conspired with the determinants in \eqref{eq:withnumerators} to produce unit numerators. In this case. the rational function associated to $\mathcal{P}$ is given by
\begin{align}
      \underline{\Omega}_{\mathcal{P}}|_{H} = \left( \sum_{v \in \mathrm{vertices}}  \prod_{\substack{f \in \mathrm{facets}  \\ v \in f }} \frac{1}{X_f} \right).
    \label{eq:rationalsubspace}
\end{align}
As for $\Omega$, we find natural to associate to $\mathcal{P}$ a rational function defined on the space of all facet variables by
\begin{align}
      \underline{\Omega}_{\mathcal{P}} = \left( \sum_{v \in \mathrm{vertices}}  \prod_{\substack{f \in \mathrm{facets}  \\ v \in f }} \frac{1}{X_f} \right),
    \label{eq:rational}
\end{align}
we reiterate that in \eqref{eq:rational} the variables $X_f$ are thought of as independent variables on an affine space.

\subsection{$\dlog$ forms and $\Res$ operator}

\begin{figure}[h!]
    \centering
    \includegraphics[width=\textwidth,trim={0cm 3.5cm 0cm 3.5cm}]{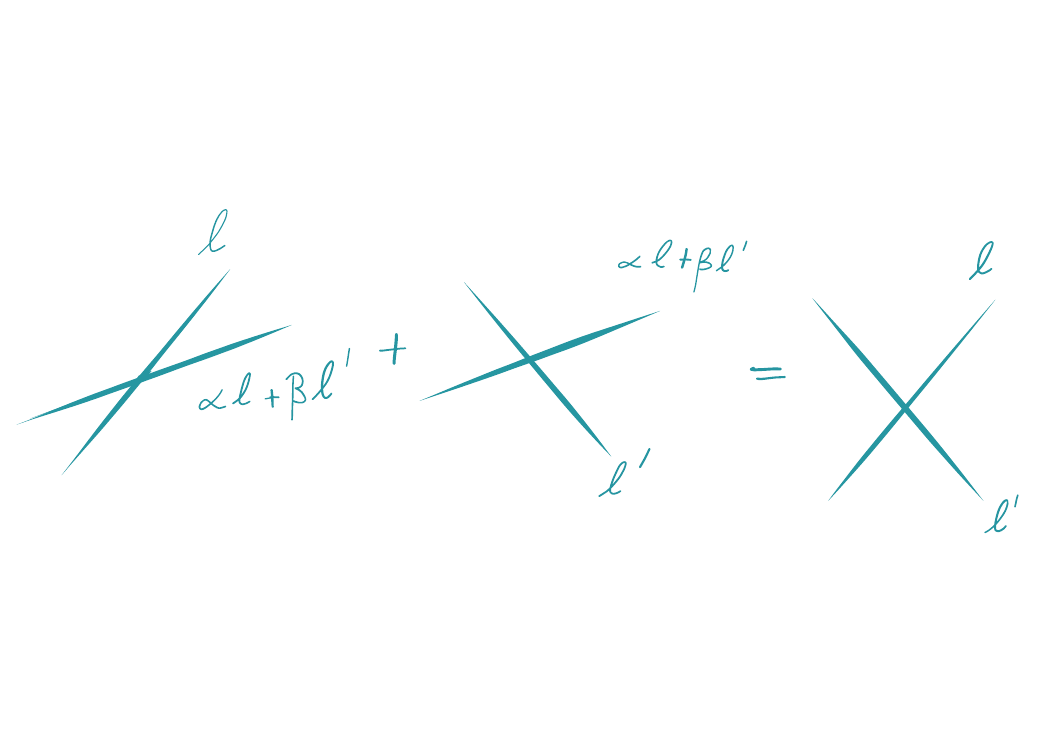}
    \caption{Geometrical interpretation of the identities between $\dlog$ forms}
    \label{fig:Trick}
\end{figure}

On the affine space $V$ with coordinates $(X_1,\dots,X_n)$ consider a family $\mathcal{A}$ of codimension one planes passing through the origin, we denote by $\ell$ the linear equations defining these planes. We will always assume that $\mathcal{A}$ contains all of the hyperplanes $\{X_i = 0\}$.
Then we define the following vector space of differential forms
\begin{align}
    R^p_V(\mathcal{A})\coloneqq\{\Omega\ |\ \Omega = 
    \sum_{\substack{\ell \subset D  \\ |\ell| = p }} c_\ell\ \dlog(\ell_1) \wedge \dots \wedge \dlog(\ell_p), c_\ell \in \mathbb{C}\}.
\end{align}
%\begin{align}
 %   \Omega^p_V(\log D)\coloneqq\{\Omega\ |\ \Omega = 
  %  \sum_{\substack{\ell \subset D  \\ |\ell| = p }} c_\ell\ \dlog(\ell_1) \wedge \dots \wedge \dlog(\ell_p), c_\ell \in \mathbb{C}\}.
%\end{align}
$R^p_V(\mathcal{A})$ is a finitely generated vector space, a set of generators is provided by the simple p-forms $\dlog(\ell_1) \wedge \dots \wedge \dlog(\ell_p)$, which are overcomplete because of partial fractions identities such as
\begin{align}
    \dlog\left(\frac{\ell}{\ell'}\right) \wedge \dlog(\alpha \ell + \beta \ell') = \dlog(\ell) \wedge \dlog(\ell') \quad \alpha,\beta \in \mathbb{C}.
    \label{eq:trick}
\end{align}
Note that \eqref{eq:trick} can be interpreted geometrically in terms of the triangulation depicted in Fig.~\ref{fig:Trick}.
A detailed study of the algebraic structure of the space $R^p_V(\mathcal{A})$ and its connection with the combinatorics of $\mathcal{A}$ can be found in \cite{orlik}. An important fact for us is that the forms
\begin{align}
    \omega_I = \dlog(X_{I_1}) \wedge \dots \wedge \dlog(X_{I_p}) \quad \mathrm{for\ any\ }I \subset (1,\dots,n),
    \label{eq:simpleform}
\end{align}
are all independent.
For any $\ell \in \mathcal{A}$, we define the residue operator as a map $$\Res_\ell : R^p_V(\mathcal{A}) \to R^{p-1}_{\ell}(\mathcal{A})$$ defined by writing any form $\Omega$ as $\Omega = \dlog{\ell}\wedge \eta + \omega$, with $\eta,\omega$ regular in $\{\ell=0\}$ and then posing $\Res_\ell \Omega = \left.\eta\right|_{\{\ell = 0\}}$. If it is not possible to write $\Omega$ in this way, then the residue is zero.

Suppose that $\Omega \in R^p_V(\mathcal{A})$ lies in the kernel of $\Res_\ell$ for some $\ell \in D$, then one can show that $\Omega \in R^p_V(\mathcal{A}')$ where $\mathcal{A}' = \mathcal{A} \setminus \{\ell\}$, in other words we can eliminate the variable $\ell$ from $\Omega$. For example, the LHS of \eqref{eq:trick} has zero residue on the plane $\{\alpha \ell + \beta \ell' =0\}$ and indeed this plane does not appear on the RHS.
However, in general the elimination of the variable may require more complicated identities, such as
\begin{align*}
    \dlog(\ell_1) \dlog(\ell_2) \dlog(\ell_3) &=\dlog(\ell_1+\ell_2+\ell_3) \wedge \nonumber\\ &\left[\dlog(\ell_1) \dlog(\ell_2) - \dlog(\ell_1) \dlog(\ell_3) + \dlog(\ell_2) \dlog(\ell_3)\right].
\end{align*}
By iterated application of this fact one can deduce that if $\Res_{\ell} \Omega = 0$ for any $\ell$, then $\Omega$ must be zero. Finally, if $\Omega$ is written as a linear combination of the forms defined in \eqref{eq:simpleform}, i.e. if
\begin{align}
\label{eq:combinationsimple}
    \Omega = \sum_{I} c_I \omega_I,
\end{align}
then we can define an iterated residue operator by $\Res_{X_J} \coloneqq \Res_{J_1} \circ  \dots \circ \Res_{J_p}$ whose action is clearly $\Res_{X_J} \Omega = c_J$. Note that $\Res_{X_J}$, when it acts on forms such as \eqref{eq:combinationsimple} is antisymmetric with respect to the planes $X_{J_i}$.

In what follows we will interested in the particularly simple case where $\mathcal{A}$ is composed by the hyperplanes $\{X_i = 0\}$, $\{X_i - X_j =0\}$ and $\{X_i + X_j = 0\}$ and $p < n$. Looking back at \eqref{eq:canonicalsimple2} and at our na\"ive guess for the canonical form of a prism \eqref{eq:naiveprism}
we see that this is the minimal choice required.

\section{The recursive formulae}
 \label{Section:formulae}
 
We are now in position to state and prove our main result, a recursive formula to compute the canonical form of a simple polytope in terms of the canonical forms of its facets. In the case of Amplituhedra facets factorize into lower dimensional Amplituhedra, mimicking the factorization of amplitudes into product of lower point amplitudes, so that the recursion for $\Omega$ yields novel, BCFW-like, expressions for the corresponding amplitudes.

\subsection{Recursive formula for $\Omega$}
\label{sec:omegarec}

Let $\mathcal{P}$ be a $d$-dimensional simple polytope, partition its facets into a distinguished one $X_b$, which can be chosen arbitrarily, and the remaining ones which we collectively denote by $\mathcal{U}$.
For each of the facet $X_u$ in $\mathcal{U}$ we introduce the operator $D_u$ that acts on differential forms by replacing $X_{u'} \to X_{u'} - X_u$, for $X_{u'} \in \mathcal{U}\setminus \{X_u\}$ and $X_b \to X_b + X_u$ \footnote{One could more generally define $D_u$ so that it sends $X \to \alpha X + \beta X_u$, but for simplicity of notation we stick to our more restrictive choice.}.
We claim that the canonical form of $\mathcal{P}$ is given by
\begin{align}
    \Omega_{\mathcal{P}} = \sum\limits_{X_u \in \mathcal{U}} \dlog\left(\frac{X_u}{X_b}\right) \wedge D_u (\pm\Omega_{X_u})
    \label{eq:recursionform}
\end{align}
where we wrote $(\pm\Omega_{X_u})$ to emphasize that the canonical forms of the facets are defined only up to an overall sign. This ambiguity is fixed by the requirement that the spurious poles along $\{X_u' - X_u=0\}$, introduced  by $D_u$, cancel each other. In order to do so, one can arbitrarily choose the sign for one of the facets in $\mathcal{U}$ and then fix the orientation of the remaining facets accordingly.
We remark that \eqref{eq:recursionform} makes manifest the projectivity of $\Omega_\mathcal{P}$, since it can be recursed to compute the canonical forms $\Omega_{X_u}$.
On the other hand, it is not obvious that the sign-fixing procedure will be consistent, therefore the cancellation of spurious poles is not manifest. This is in complete analogy with \eqref{eq:canonicalsimple2}, where no spurious poles were introduced, but the sign-fixing procedure to ensure projectivity was not obviously consistent. 

Let us now prove the correctness of \eqref{eq:recursionform}, by showing that it satisfies the definition \eqref{eq:canonicalDef}. We assume that signs have been chosen so that $\Res_{(X_u' - X_u)} \Omega_{\mathcal{P}} = 0$. We also have spurious poles at $\{X_u + X_b = 0\}$, but their cancellation is clear due to the prefactor $\dlog\left(\frac{X_u}{X_b}\right)$. From the absence of spurious poles follows, in particular, that $\Omega_{\mathcal{P}}$ is written in terms of $\dlog$s involving only the facet variables $X$ and thus any iterated residue $\Res_{X_J}(\Omega_{\mathcal{P}})$ is anti-symmetric with respect to the facet variables $X_{J_i}$. The pole in $\{X_u = 0\}$ appear in \eqref{eq:recursionform} only through the prefactor $\dlog\left(\frac{X_u}{X_b}\right)$ and since evaluating the residue there undo the action of the operator $D_u$, we have $\Res_{X_u} \Omega_{\mathcal{P}} = \Omega_{X_u}$.
It is left to check that $\Res_{X_b} \Omega_{\mathcal{P}} = \Omega_{X_b}$, applying the definition $\eqref{eq:canonicalDef}$ to $\Omega_{X_b}$ this is equivalent to $\Res_{X_u} \Res_{X_b}\Omega_{\mathcal{P}} = \Omega_{X_u \cap X_b}$ or zero if $X_u \cap X_b = \emptyset$. Since iterated residues are anti-symmetric we can invert the order of the residues and then the result follows from $\Res_{X_u} \Omega_{\mathcal{P}} = \Omega_{X_u}$.

As it stands \eqref{eq:recursionform} is a recursion formula that requires as input the knowledge of the canonical forms of all but one the facets. It turns out that the recursion can be made dramatically more efficient. After a distinguished facet $X_b$ is chosen, let us now further partition the remaining facets in two sets $\mathcal{U}$ and $\mathcal{S}$ in such a way that every vertex of $\mathcal{P}$ lies in at least one of the facets $\{X_b\} \cup \mathcal{U}$.
Surprisingly, $\eqref{eq:recursionform}$ gives again the correct canonical form. 
However, first one has to decide how to fix the relative signs of the forms $\Omega_{X_u}$, since now the $X_u$ facets might not be joined to each other by a sequence of adjacencies. In order to overcome this problem, one has to further require that if two facets $X_u$ and $X_{u'}$ are connected by a one-dimensional edge, then $\Omega_{X_u}$ and $\Omega_{X_u'}$ must have opposite residues at the vertices of that edge.
The only new element in the proof of \eqref{eq:recursionform} is that now one has to consider residues of the form \begin{align}
    \Res_{X_S,X_b}\Omega_{\mathcal{P}} = -\sum_{X_u \in \mathcal{U}} \Res_{X_S} D_u\Omega_{X_u} = -\sum_{X_u \in\mathcal{U}} \Res_{X_S} \Omega_{X_u},
    \label{eq:newresidue}
\end{align} 
where $X_S$ is a (d-1)-tuple of facets in $\mathcal{S}$ and in the second passage we used the fact that $D_u$ does not deform the variables $\mathcal{S}$, so that $\Res_{X_s} D_u = D_u \Res_{X_s}$ for any pair $s \in \mathcal{S}$ and $u \in \mathcal{U}$. Let us write $\cap X_S$ for the intersection of the corresponding facets, then there are three cases to be considered: $\cap X_S$ is empty, $\cap X_S$ defines a one dimensional edge of $\mathcal{P}$ which is incident to two facets in $\mathcal{U}$ and $\cap X_S$ defines a one dimensional edge incident to a facet $X_u$ in $\mathcal{U}$ and to the facet $X_b$. 
In the first case, each of the terms in the RHS of \eqref{eq:newresidue} is zero, in the second case the sum is zero and in the third case \eqref{eq:newresidue} gives $\Res_{X_S,X_b}\Omega_{\mathcal{P}}=- \Res_{X_S,X_u}\Omega_{\mathcal{P}}$. In every case, the result is correct.

\begin{figure}[h!]
    \centering
    \includegraphics[width=.5\textwidth,trim={15cm 10cm 15cm 25cm}]{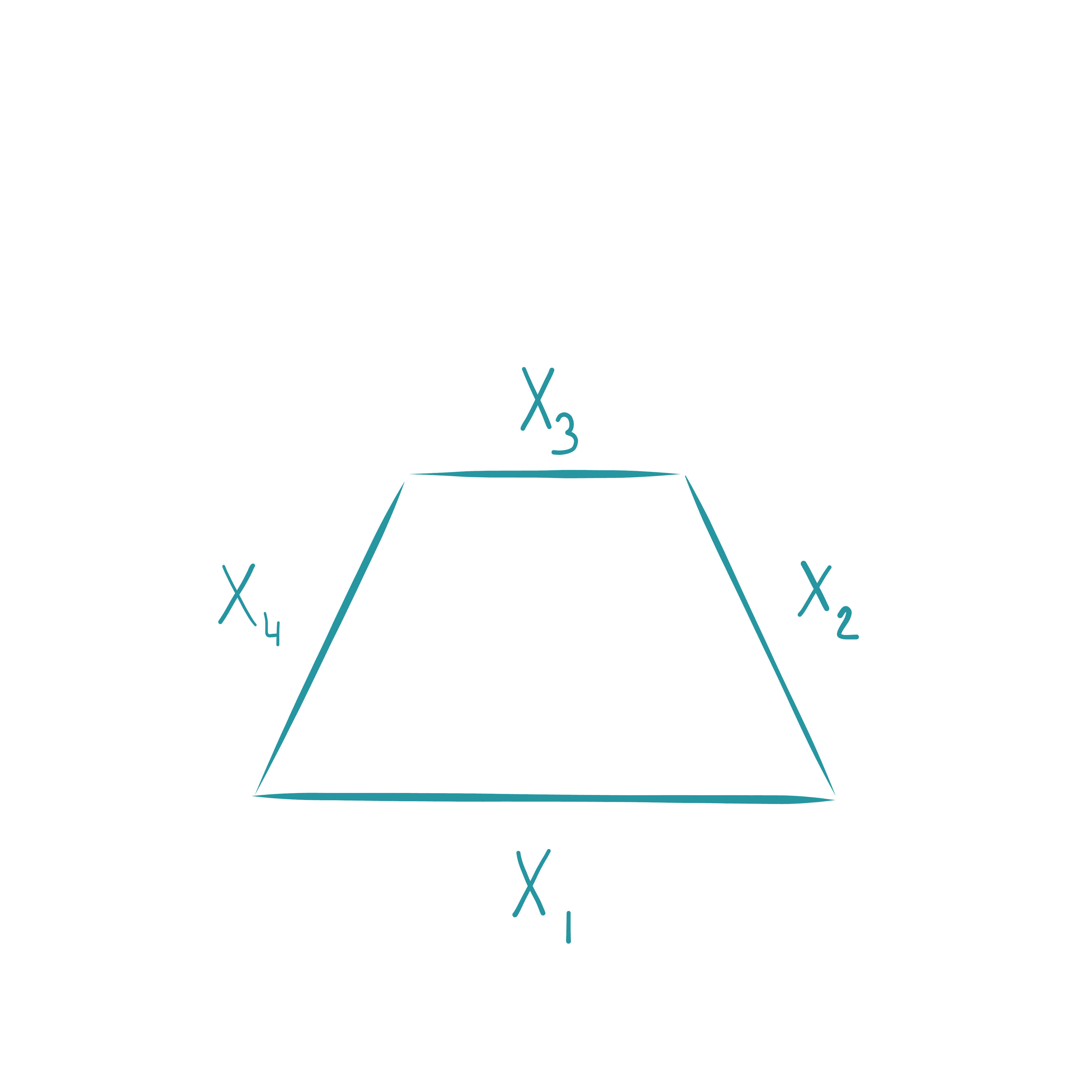}
    \caption{A square}
    \label{fig:square}
\end{figure}

We conclude this section with a few explicit examples of \eqref{eq:recursionform}.
The simplest example is a segment with facets $X_1$ and $X_2$, and we trivially get $\Omega = \dlog\left(\frac{X_1}{X_2}\right).$
Already the case of a 2-simplex with facets $\{X_1,X_2,X_3\}$ is slightly interesting,  we can choose either $X_b = X_1$, $\mathcal{U}=\{X_2,X_3\}$ and $\mathcal{S}$ empty or $\mathcal{U}=\{X_2\}$ and $\mathcal{S}=\{X_3\}.$ We obtain, respectively,
\begin{align*}
 \Omega = \dlog\left(\frac{X_2}{X_1}\right)\wedge  \dlog\left(\frac{X_3-X_2}{X_2+X_1}\right)  - \dlog\left(\frac{X_3}{X_1}\right)\wedge  \dlog\left(\frac{X_2-X_3}{X_3+X_1}\right),
\end{align*}
and
\begin{align*}
 \Omega = \dlog\left(\frac{X_2}{X_1}\right)\wedge  \dlog\left(\frac{X_3}{X_2+X_1}\right) ,
\end{align*}
using \eqref{eq:trick} it is easy to see that both agree with \eqref{eq:canonicalsimple2}, which would give
\begin{align}
    \Omega = \dlog{X_1}\wedge\dlog{X_2}-\dlog{X_3}\wedge\dlog{X_2} - \dlog{X_1}\wedge\dlog{X_3}.
\end{align}
A more interesting example is provided by a two dimensional square with facets\\ $\{X_1,X_2,X_3,X_4\}$, see Fig. \ref{fig:square}.
We consider $X_b = X_1$, and either $\mathcal{U}=\{X_3\}$ or $\mathcal{U}=\{X_2,X_4\}.$
In the first case we get
\begin{align}
    \Omega = \dlog\left(\frac{X_3}{X_1}\right) \wedge \dlog\left(\frac{X_2}{X_4}\right),
\end{align}
which is clearly the form of the square $[X_1,X_3] \times [X_2,X_4]$.
In the other case we get
\begin{align}
    \Omega_{\pm} = \dlog\left(\frac{X_2}{X_1}\right) \wedge \dlog\left(\frac{X_3}{X_1+X_2}\right)\pm \dlog\left(\frac{X_4}{X_1}\right) \wedge \dlog\left(\frac{X_3}{X_1+X_4}\right),
\end{align}
we wish to emphasize that, regardless of the relative signs we choose, $\Omega_{\pm}$ does not have spurious poles along $\{X_{4}-X_{2}=0\}$, but $\Res_{X_2,X_1}\Omega_{+}=\Res_{X_4,X_1}\Omega_{+}=1$, so that we have to pick $\Omega_{-}$ to satisfy the sign fixing rule.

\subsection{Recursive formula for \underline{$\Omega$}}

We recall the expression \eqref{eq:rational} for the canonical rational function $\underline{\Omega}_{\mathcal{P}}$ associated to a simple polytope $\mathcal{P}$,
\begin{align}
      \underline{\Omega}_{\mathcal{P}} = \left( \sum_{v \in \mathrm{vertices}}  \prod_{\substack{f \in \mathrm{facets}  \\ v \in f }} \frac{1}{X_f} \right).
      \label{eq:sumvertices}
\end{align}
Suppose that the facets of $\mathcal{P}$ are partitioned in $X_b$ and two sets $\mathcal{U}, \mathcal{S}$, as described in the previous section. Staring at the recursive formula $\eqref{eq:recursionform}$ for $\Omega$ one would na\"ively guess that a similar expression should exist for $\underline{\Omega}_{\mathcal{P}}$, since after all the proof of $\eqref{eq:recursionform}$ relies on partial fraction identities, such as $\eqref{eq:trick}$, which also hold at the level of functions. We consider, then, the following expression
\begin{align}
\label{eq:rationalrec}
    \underline{\Omega} = \sum_{X_u \in \, \mathcal{U}} \left( \frac{1}{X_b} + \frac{1}{X_{u}} \right)D_{u} \underline{\Omega}_{u}
\end{align}
where the deformation operators $D_{u}$ are again defined by $X_{u'} \to X_{u'} - X_{u}$ for $u \ne u'$ and $X_b \to X_b + X_u$. Note that differential forms carry a notion of an orientation, which is crucial in the cancellation of spurious poles in $\Omega$. At the level of $\underline{\Omega}$ this notion is translated in the positivity of the functions appearing in \eqref{eq:rationalrec}, that is in the fact that $D_u X_u' = - D_u' X_u$.
It turns out, however, that $\eqref{eq:rationalrec}$ holds only if the choice of $\mathcal{U}$ and $\mathcal{S}$ is such that there is no $(d-1)$-tuple of facets $X_s \in \mathcal{S}$ whose intersection defines a one dimensional edge of $\mathcal{P}$ adjacent to two facets in $\mathcal{U}$. 
This should not be suprising thinking back at the proof of  $\eqref{eq:recursionform}$, the existence of such an edge requires to orient the forms $\Omega_{X_u}$ so that they have opposite residues at its vertices, a requirement which is not easily translated at the level of $\underline{\Omega}$.

Before proving \eqref{eq:rationalrec} in full generality, let us give an example to better illustrate both the formula and the aforementioned extra condition.
We will consider once again the square of Fig. \ref{fig:square}, with the same choices of $\mathcal{U}$ and $\mathcal{B}$.
The canonical form of the square is given by \begin{align}
    \underline{\Omega} = \frac{1}{X_1 X_2}+\frac{1}{X_2 X_3}+\frac{1}{X_3 X_4}+\frac{1}{X_1 X_4}.
\end{align}
In the case where $\mathcal{U} = \{X_3\}$ and $\mathcal{S}$ is empty the recursive formula reads
\begin{align}
    \underline{\Omega} = \left(\frac{1}{X_1}+\frac{1}{X_3}\right)\left(\frac{1}{X_2} + \frac{1}{X_4}\right),
\end{align}
while if we chose $\mathcal{U} = \{X_2,X_4\}$ and $\mathcal{S} = \{X_3\}$ we would \emph{incorrectly} get
\begin{align}
    \underline{\Omega} = \left(\frac{1}{X_1}+\frac{1}{X_2}\right)\left(\frac{1}{X_1+X_2} + \frac{1}{X_3}\right)+
    \left(\frac{1}{X_1}+\frac{1}{X_4}\right)\left(\frac{1}{X_1+X_4} + \frac{1}{X_3}\right),
\end{align}
the mismatch is due to the contributions from the vertices on the facet $\{X_3=0\}$.
Indeed, this choice of $\mathcal{U}$ and $\mathcal{S}$ does not satisfy the new requirement because the one dimensional edge $X_3 \in \mathcal{S}$ is incident to both facets in $\mathcal{U}$.

We now prove that $\underline{\Omega}_{\mathcal{P}}$ as defined by $\eref{eq:sumvertices}$ is the same as computed in $\eref{eq:rationalrec}$.
Our strategy will be to consider the contribution of every vertex $v$ in $\mathcal{P}$ to both expressions. There are several types of vertices to consider:
\begin{enumerate}
    \item Vertex $v \notin X_b$ belongs to only one of the elements of $\mathcal{U}$. Then, there is only one term in (\ref{eq:rationalrec}) to consider, of the form
    
    \begin{align}
        \left( \frac{1}{X_b} + \frac{1}{X_{u_i}} \right) \frac{1}{X_{s_1} \hdots X_{s_{d-1}}} = \frac{1}{X_b X_{s_1} \hdots X_{s_{d-1}}} + \frac{1}{X_{u_i} X_{s_1} \hdots X_{s_{d-1}}}
    \end{align}
    
    reproducing both the term in \eref{eq:sumvertices} corresponding to $v$ and the term in \eref{eq:sumvertices} corresponding to $B=\bigcap_{a=1}^{d-1} X_{s_i} \cap X_b$. Since $d-1$ facets in $\mathcal{S}$ can not intersect along an edge adjacent to two vertices in $\mathcal{U}$, $B$ will be one of the vertices of $\mathcal{P}$, and it is easy to check that its corresponding term will not appear again in \eref{eq:rationalrec}. 
    
    \item Vertex $v \notin X_b$ belongs to a collection of $n \geq 2$ facets in $\mathcal{U}$, $X_{u_1}$, $X_{u_2}$, $\hdots$ ,$X_{u_n}$. Then, we need to consider the sum of $n$ terms in \eref{eq:rationalrec} of the form 
    
    \begin{align}
    \label{eq:case2}
        \sum_{i=1}^{n} \left( \frac{1}{X_b} + \frac{1}{X_{u_i}} \right) \prod_{\substack{j = 1 \\ j \neq i}}^n \left( \frac{1}{X_{u_j}} - \frac{1}{X_{u_i}} \right) \frac{1}{X_{s_1} \hdots X_{s_{d-n}}}
    \end{align}
    
    Now, we can use the following two identities,
    
    \begin{align}
    \label{eq:partialfrac}
        \sum_{i=1}^{n} \frac{1}{X_{u_{i}}} \prod_{\substack{j = 1 \\ j \neq i}}^n& \frac{1}{X_{u_{j}} - X_{u_{i}}} = \frac{1}{X_{u_1} X_{u_2} \hdots X_{u_n}} \\
    \label{eq:partialfrac2}
        &\sum_{i=1}^{n} \prod_{\substack{j = 1 \\ j \neq i}}^n \frac{1}{X_{u_j} - X_{u_i}} = 0
    \end{align}
    
    to conclude that the sum from \eref{eq:case2} will reproduce the term corresponding to $v$ in \eref{eq:sumvertices}.
    
    \item Vertex $v \in X_b$ is shared with only one element of $\mathcal{U}$, which we will denote as $X_{u_i}$. Then, there will only be one term in \eref{eq:rationalrec} to consider,
    
    \begin{align}
        \left( \frac{1}{X_b} + \frac{1}{X_{u_i}} \right) \frac{1}{(X_b + X_{u_i}) X_{s_1} \hdots X_{s_{d-2}} } = \frac{1}{X_b X_{u_i} X_{s_1} \hdots X_{s_{d-2}} }
    \end{align}
    
    reproducing the term corresponding to $v$ in \eref{eq:sumvertices}.
    
    \item Vertex $v \in X_b$ is shared with a collection of $n$ elements of $\mathcal{U}$, which we will denote as $X_{u_1}$, $X_{u_2}$, $\hdots$ , $X_{u_n}$. Then, there are $n$ terms of \eref{eq:rationalrec} to consider,
    
    \begin{align}
        \sum_{i=1}^{n} \left( \frac{1}{X_b} + \frac{1}{X_{u_i}} \right) \prod_{\substack{j = 1 \\ j \neq i}}^n \left( \frac{1}{X_{u_j} - X_{u_i}} \right)
        \frac{1}{X_b + X_{u_i}}
        \frac{1}{X_{s_1} \hdots X_{s_{d-n-1}}} = \nonumber \\
        \sum_{i=1}^{n}  \frac{1}{X_{u_i}} \prod_{\substack{j = 1 \\ j \neq i}}^n \left( \frac{1}{X_{u_j} - X_{u_i}} \right)
        \frac{1}{X_b X_{s_1} \hdots X_{s_{d-n-1}}} = \frac{1}{X_b X_{u_1} X_{u_2} \hdots X_{u_n} X_{s_1} \hdots X_{s_{d-n-1}}}
    \end{align}
    
    where the identity in \eref{eq:partialfrac} was used in the second term. We can see that the result matches the appropriate term of \eref{eq:sumvertices}. 
    
\end{enumerate}

\section{Application to planar $\phi^4$ amplitudes}
\label{Section:phi4}

In the context of planar $\phi^4$ amplitudes, Feynman diagrams for $n$ - particle amplitudes correspond to the quadrangulations of an $n$ - gon. Unlike the bi-adjoint $\phi^3$ theory, a single polytope whose vertices are in correspondence with the full set of quadrangulations does not exist. Instead, a notion of \textit{compatibility} with a reference quadrangulation $Q$ is used to select a subset of all quadrangulations which correspond to a polytope, called Stokes polytope. The full amplitude is then obtained by summing the contributions coming from all Stokes polytope, each weighted by an appropriate factor that takes into account overcounting of individual Feynman diagrams, further details can be found in \cite{Baryshnikov}.

The properties of Stokes polytopes have been studied by mathematicians in \cite{Chapoton},\cite{PaluPilaud}, and used by physicists to describe the planar $\phi^4$ amplitudes in \cite{Banerjee:2018tun}, \cite{Aneesh:2019cvt}. For the sake of self-containedness we offer in the rest of this section a quick review of the most important facts concerning Stokes polytopes while we refer to the literature for more details and proofs.
 
\begin{figure}
\centering
\includegraphics[width=0.4\textwidth]{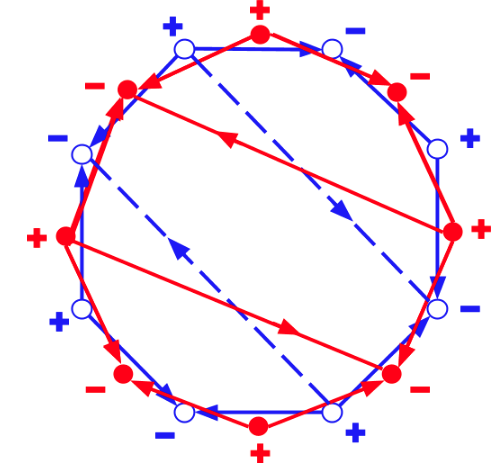}
\caption{\label{fig:qcomp} An example of a red quadrangulation compatible with a blue one}
\end{figure}

We start by defining Q-compatibility. Choose an alternating assignment of `+' and `-' to the vertices of the polygon \footnote{We will choose an assignment of '+' to odd and '-' to even - numbered vertices} and assign an arrow to each diagonal, pointing from a `+' vertex to a `-' one. Then, we consider a slight clockwise rotation of our polygon superimposed onto itself, as in Fig.~\ref{fig:qcomp}. Draw the reference quadrangulation $Q$ in blue, and draw in red any quadrangulation $Q'$ of the slightly rotated polygon. We say that $Q'$ is Q-compatible if for any pair of a blue and a red crossing arrows the rotation with the smallest angle that makes the blue arrow to point in the same direction as the red arrow is clockwise.
 
All quadrangulations $Q'$ compatible with $Q$ correspond to vertices of a Stokes polytope $S_Q$, which is a simple polytope.
There is a ``mutation rule'' allowing to move from any vertex to an adjacent one, thus exploring the edges of the polytope, which is a generalization of a similar rule in the case of triangulations of an n-gon: removing any diagonal of a quadrangulation $Q'$ produces an hexagon which can be quadrangulated in three different ways by adding a diagonal. However, only two of the resulting quadrangulations will be Q-compatible, these corresponds to two adjacent vertices in $S_Q$.
Let us warn the reader that Q-compatibility is not an equivalence relation, in other words the fact that a quadrangulation $Q'$ appears among the vertices of $S_Q$ does not imply that $Q$ appears among the vertices of $S_{Q'}$.
Unlike the case of triangulations, there is not a single polytope associated to an n-gon. In fact, it turns out that for $2 n + 2$ - gon, there are $\frac{1}{2 n + 1} {{3 n}\choose{n}}$ distinct Stokes polytopes, one for each of the quadrangulations of the $2 n + 2$ - gon taken as a reference.

Stokes polytopes are simple polytopes with the property that each facet $X_{i j}$ factorizes as $\partial_{X_{i j}} S_Q = S_{Q'} \times S_{Q''}$, for some suitably chosen reference quadrangulations $Q'$, $Q''$  \cite{Manneville:2018}. Due to this recursive structure they are a natural playground where to apply our recursive formulae, which will be done in the remaining part of this section.
In order to keep track both of the dependence on the reference quadrangulation Q and of the factorization properties of Stokes polytopes and of their canonical forms, we find convenient to adopt the notation of writing
\begin{align}
\label{eq:stokescanonical}
    \Omega \addpicwide{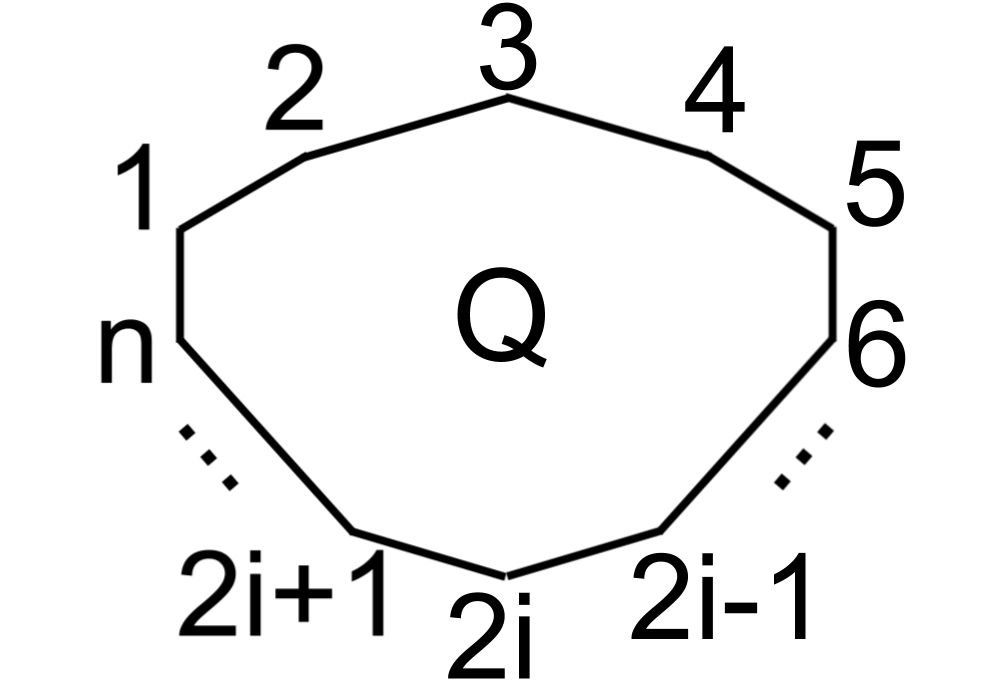},
\end{align}
for the canonical form of the Stokes polytope $S_Q$, and similarly for the rational function $\underline{\Omega}$.

\subsection{Choosing the $\mathcal{U}$ facets}

We start by noting that any Stokes polytope $S_Q$ has a face corresponding to the diagonal of the $n$ - gon of the form $(i, i+3)$. For illustration purposes, let us take this to be our face $X_b$. Then, we will define $\mathcal{U}$ to be the set of faces of $S_Q$ whose corresponding diagonals cross $(i,i+3)$. It is then easy to check that every quadrangulation will contain either the quadrilateral $(i,i+1,i+2,i+3)$ or a diagonal crossing $(i,i+3)$, hence every vertex will be contained in either one of the $\mathcal{U}$ facets or $X_b$. We may also note that no element of thus chosen $\mathcal{U}$ has intersection with $X_b$.

\begin{figure}
    \centering
    \includegraphics[width=0.4 \textwidth]{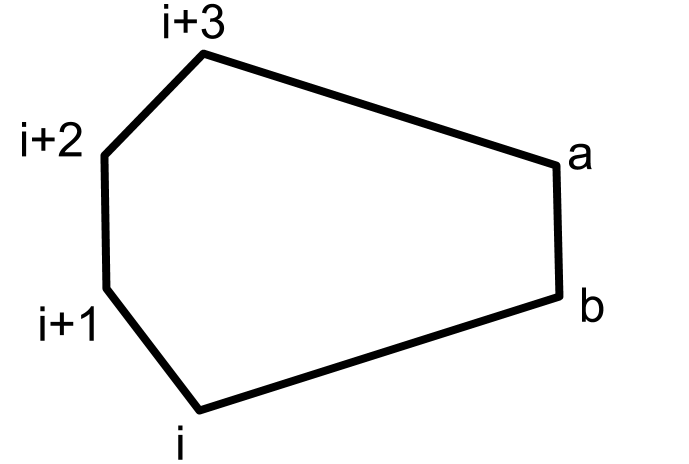}
    \caption{The hexagon carved out by a collection of diagonals corresponding to facets in $\mathcal{S}$}
    \label{fig:simpleproof}
\end{figure}

As usual, we will denote by $\mathcal{S}$ the complement of $\mathcal{U} \cup \{ X_b \}$. In order to be able to use our choice of $X_b$ and $\mathcal{U}$ to recursively compute the canonical function, we also need to check that no edge adjacent to two facets in $\mathcal{U}$ can be given as an intersection of $d-1$ facets in $\mathcal{S}$. The facets $\mathcal{S}$ correspond to Q-compatible diagonals other than $(i,i+3)$ which do not cross $(i,i+3)$. For a $d$ - dimensional $S_Q$, any subset of $d-1$ such elements of $\mathcal{S}$ will either contain a pair of crossing diagonals, in which case the intersection of such faces will be an empty set, or will form a partial quadrangulation of the form shown in Fig.~\ref{fig:simpleproof}. It is now easy to see that this quadrangulation can be completed by either adding the diagonal $(i,i+3)$ (corresponding to $X_b$), or one of the diagonals $(i+1,a)$, $(i+2,b)$ (corresponding to one of the facets in $\mathcal{U}$) selected by the Q-compatibility, and the desired property is satisfied. 

\subsection{Recursion for \underline{$\Omega$}$_{S_{Q}}$}

Note that, up to cyclic shifts of the number labels, any reference quadrangulation $Q$ will contain a facet of the form $X_{1, 4}$. Let us pick $X_b = X_{1,4}$ and choose the $\mathcal{U}$ facets as those corresponding to $Q$ - compatible variables crossing $X_{1,4}$. It is easy to check that all $\mathcal{U}$ variables must be of the form $X_{3, 2i}$ and the recursive formula for canonical function of $\mathcal{S}_Q$ takes the form

\begin{align}
\label{eq:stokesrecursion}
    \underline{\Omega} \addpicwide{Q.png} = \sum_i \left( \frac{1}{X_{1,4}} + \frac{1}{X_{3,2 i}}  \right) D_{X_{3,2i}}\underline{\Omega} \addpic{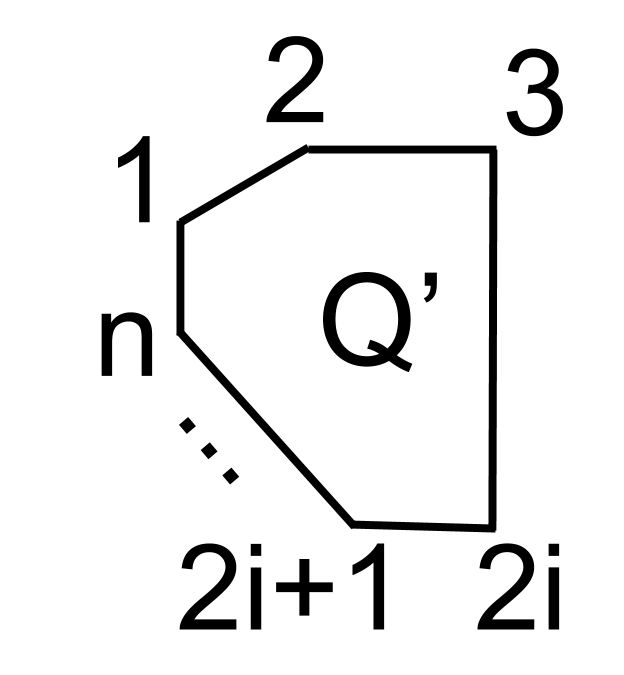} \underline{\Omega} \addpic{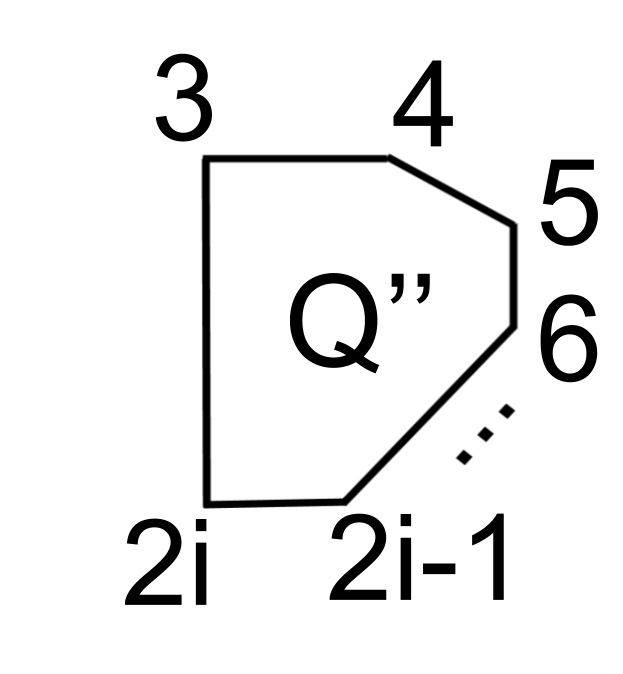}
\end{align}

where we are summing over all Q-compatible $X_{3,2i}$. The new reference quadrangulations $Q'$, $Q''$ are chosen by keeping the diagonals of $Q$ not crossing $X_{3, 2 i}$ and choosing their completion in the two regions such as to obtain the top quadrangulation corresponding to facet $X_{3,2i}$ of the oriented flip graph\footnote{We thank F.Chapoton for explaining how factorisation can be seen from the oriented flip graphs}, as defined in \cite{Chapoton}. We have verified that this produces correct results for $n \leq 10$ particles. 

 Since the facet $X_b$ only enters  \eref{eq:stokesrecursion} through the prefactors, it is particularly simple to take the limit $X_b \rightarrow \infty$, a fact we will exploit as we piece together the scattering amplitude from the limits of Stokes polytopes.

\subsection{Examples}

The canonical function of a Stokes polytope corresponding to an $n = 4$ particle amplitude is trivial and can form the basis of the recursion,

\begin{align}
    \underline{\Omega} \addpic{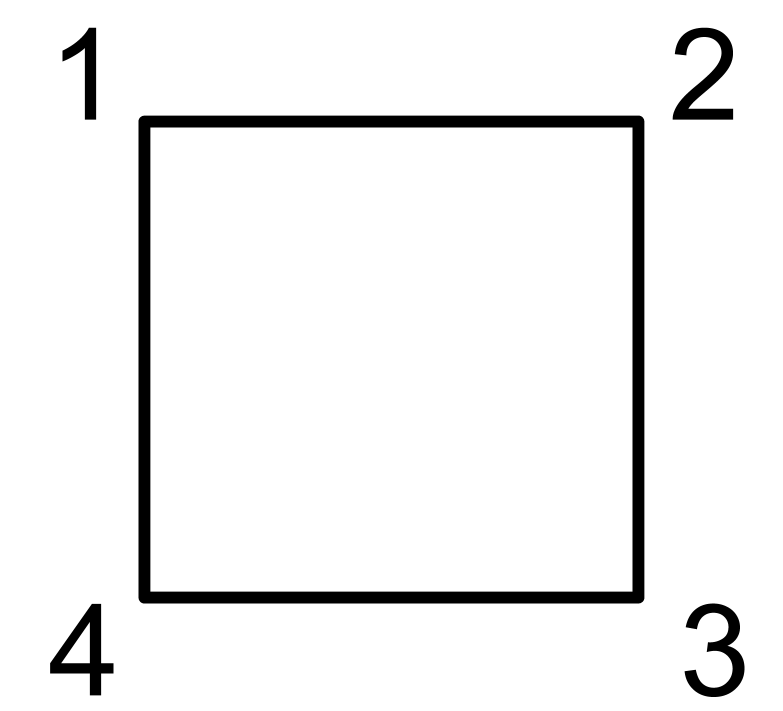} = 1
\end{align}

For the case of quadrangulations of a hexagon, it is straightforward to use the recursive formula to get the correct answer, as in the following example,

\begin{align}
    \underline{\Omega} \addpic{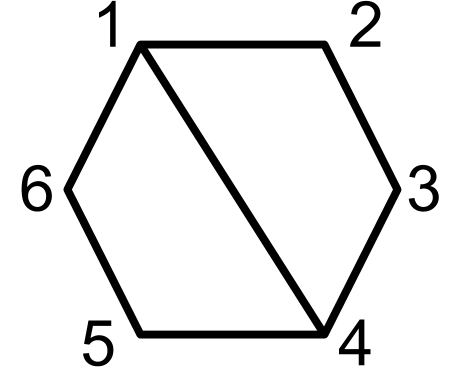} &= \left( \frac{1}{X_{1,4}} + \frac{1}{X_{3,6}} \right)  D_{X_{3,6}} \underline{\Omega} \addpic{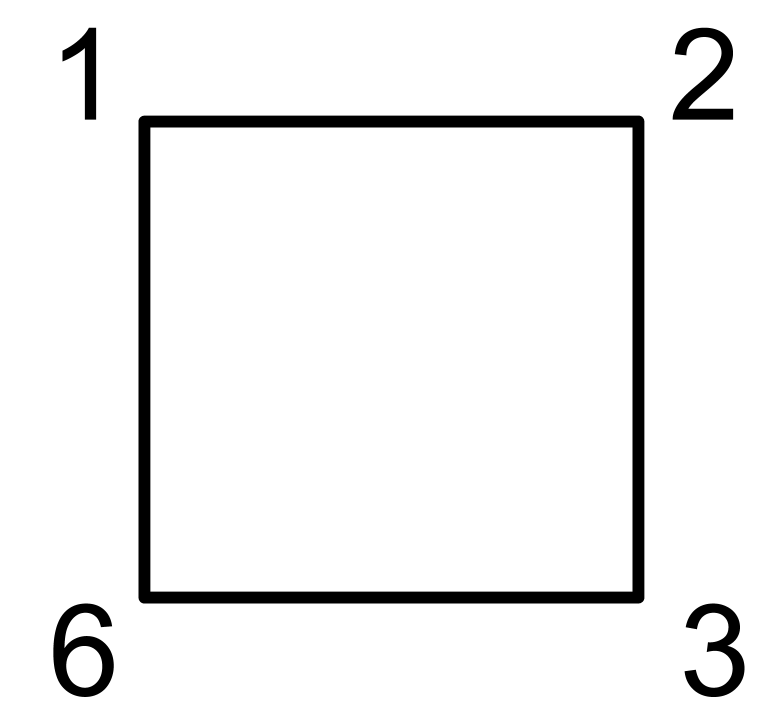} D_{X_{3,6}} \underline{\Omega} \addpic{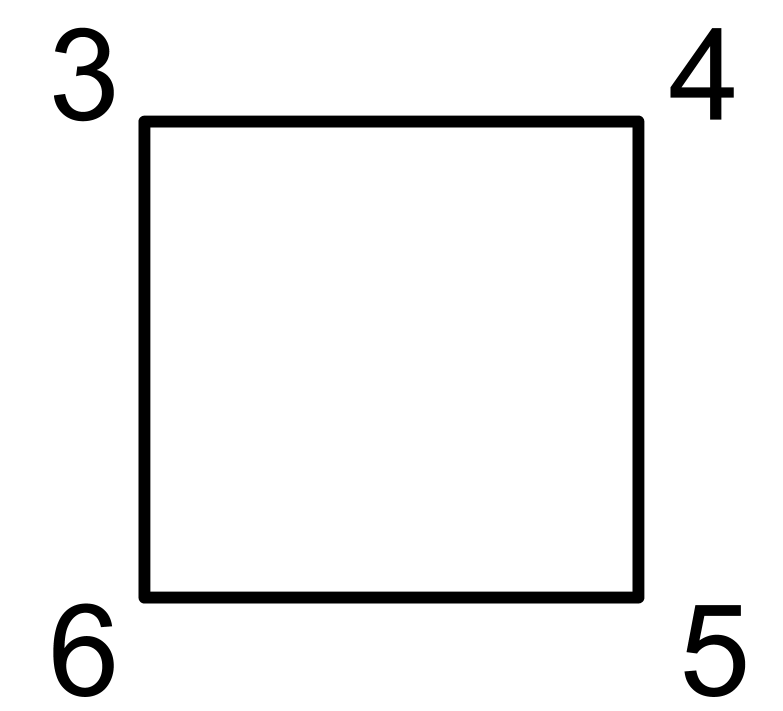} 
    \nonumber \\
    &= \frac{1}{X_{1,4}} + \frac{1}{X_{3,6}}
\end{align}

The set of quadrangulations of octagon has 12 elements corresponding to vertices of the two different types of Stokes polytopes, a box and a pentagon \cite{Banerjee:2018tun}. The canonical function of a representative of the box class is given by

\begin{align}
\label{eq:8ptbox}
     \underline{\Omega}\addpic{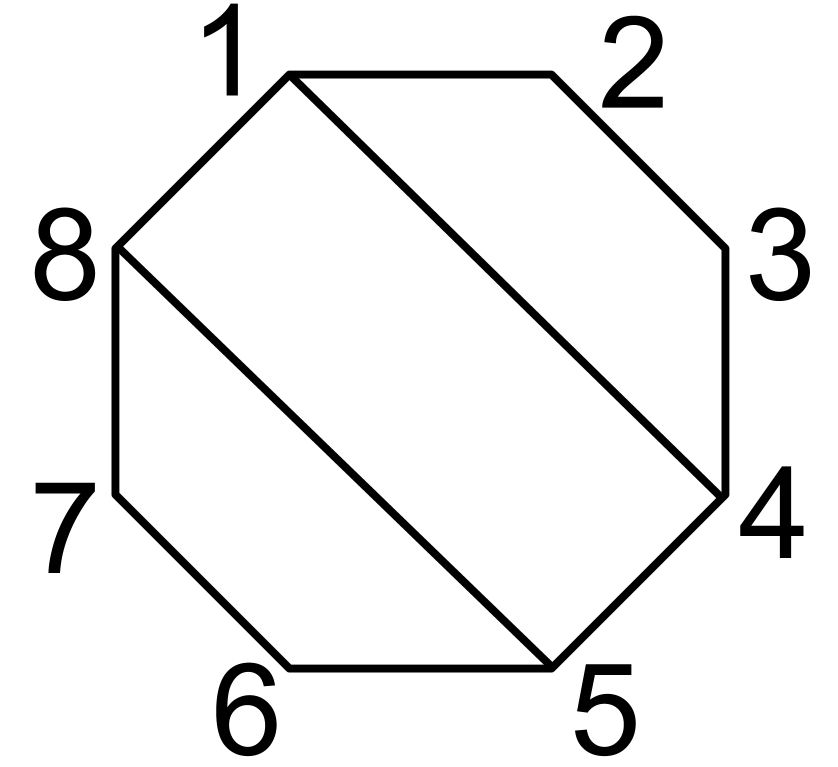} &=  \left( \frac{1}{X_{1,4}} + \frac{1}{X_{3,8}} \right) 
   D_{X_{3,8}} \underline{\Omega} \addpic{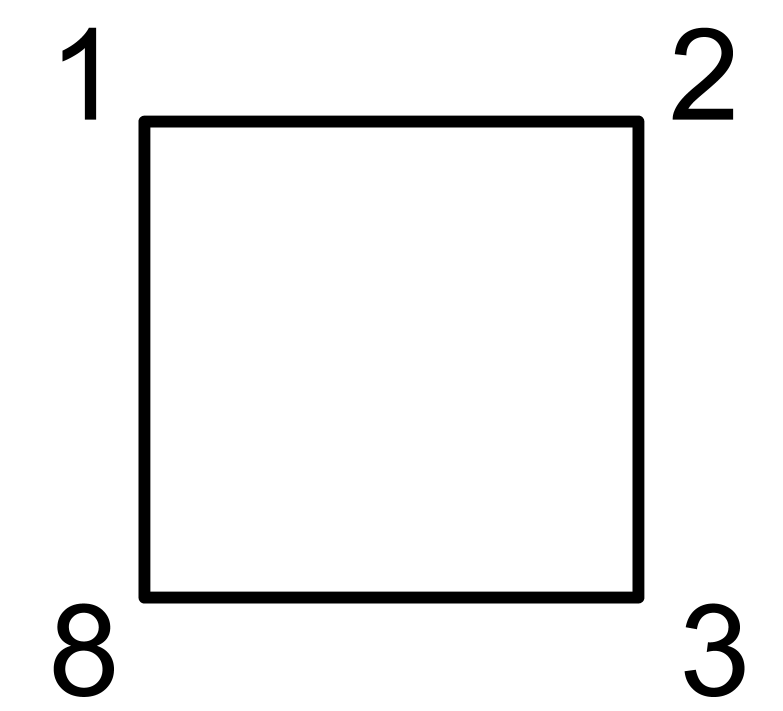} 
   D_{X_{3,8}} \underline{\Omega} \addpic{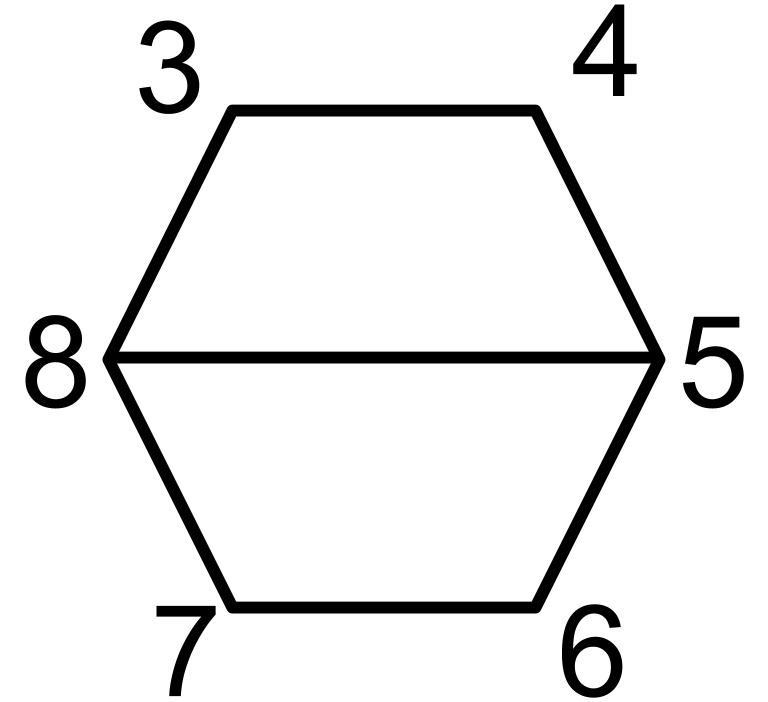} 
   \nonumber \\
  & = \left( \frac{1}{X_{1,4}} + \frac{1}{X_{3,8}} \right) 
   \left( \frac{1}{X_{4,7}} + \frac{1}{X_{5,8}}  \right) 
\end{align}

The canonical function of a representative pentagon is given by

\begin{align}
\label{eq:8ptpentagon}
   \underline{\Omega}\addpic{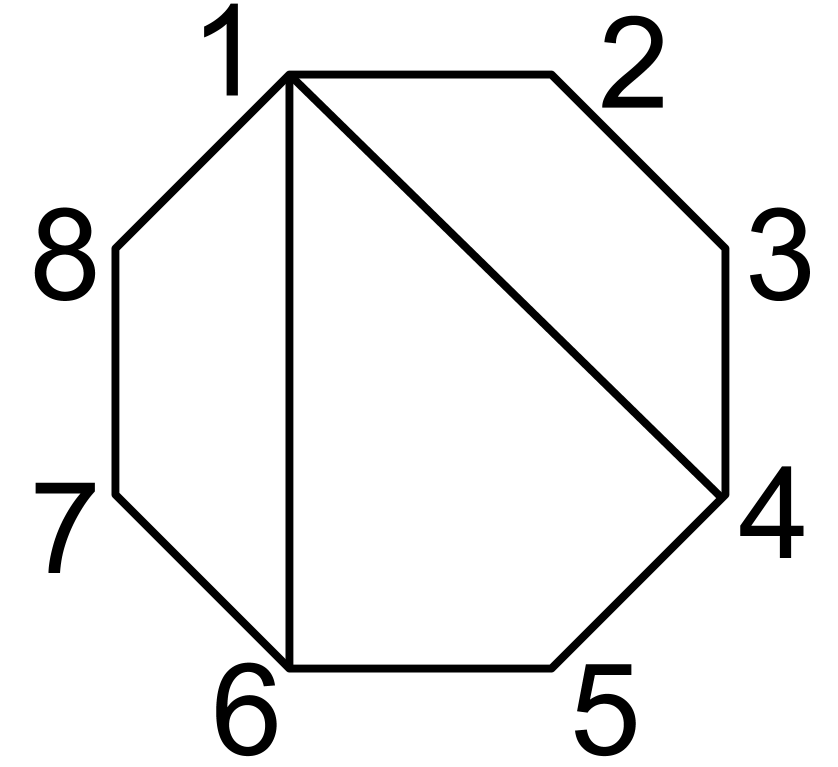} &= \left( \frac{1}{X_{1,4}} + \frac{1}{X_{3,6}} \right) D_{X_{3,6}} \underline{\Omega}  \addpic{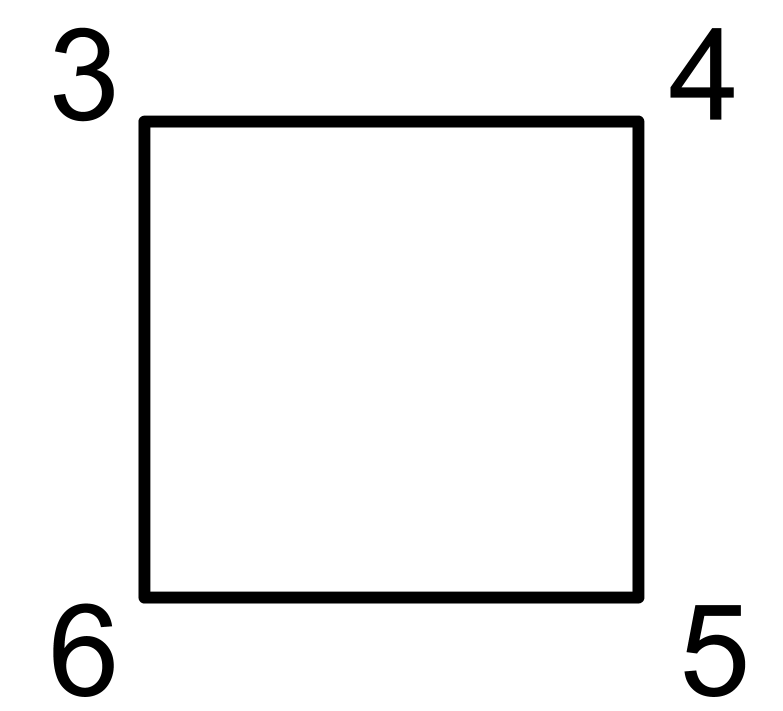} 
   D_{X_{3,6}} \underline{\Omega}  \addpic{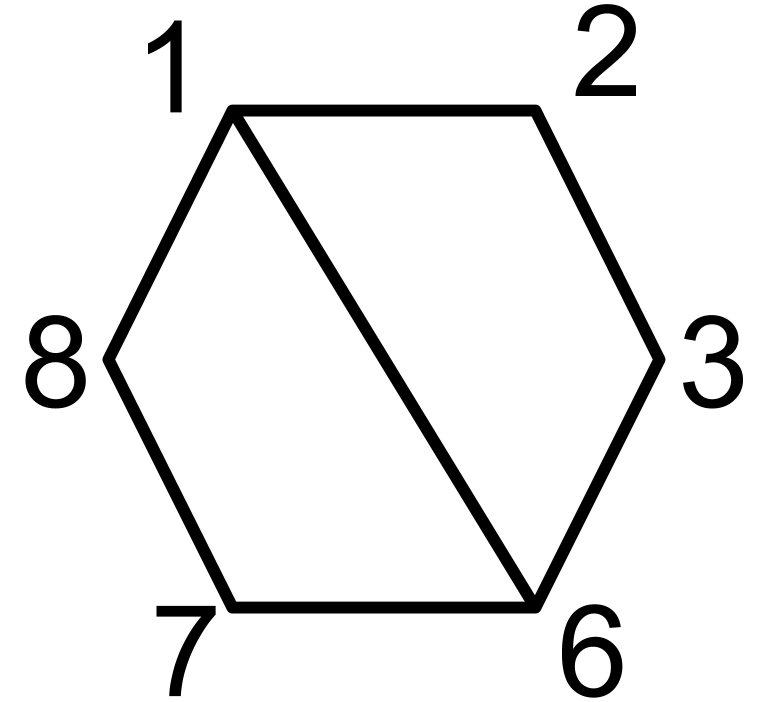} + \nonumber \\
   &+ \left( \frac{1}{X_{1,4}} + \frac{1}{X_{3,8}} \right) D_{X_{3,8}} \underline{\Omega} \addpic{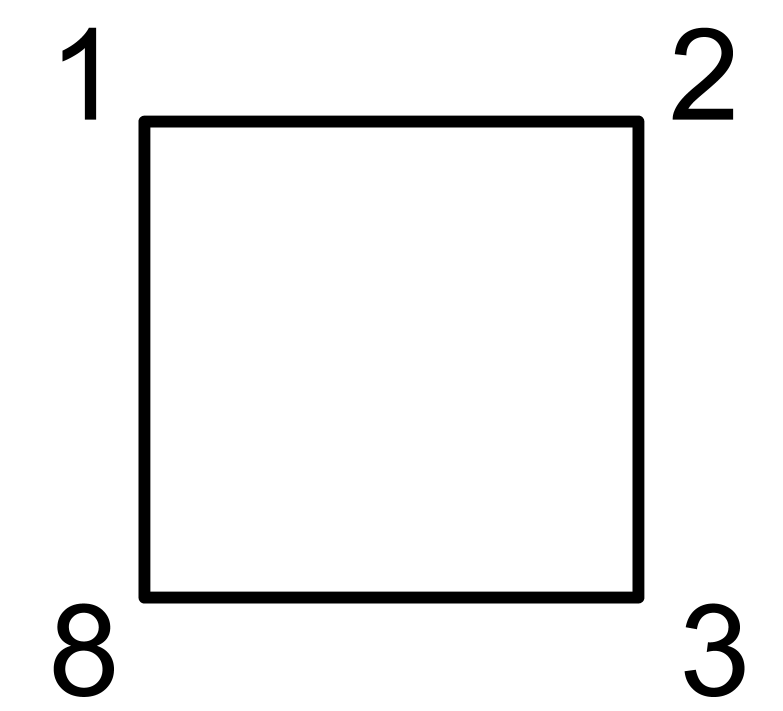} 
   D_{X_{3,8}}\underline{\Omega} \addpic{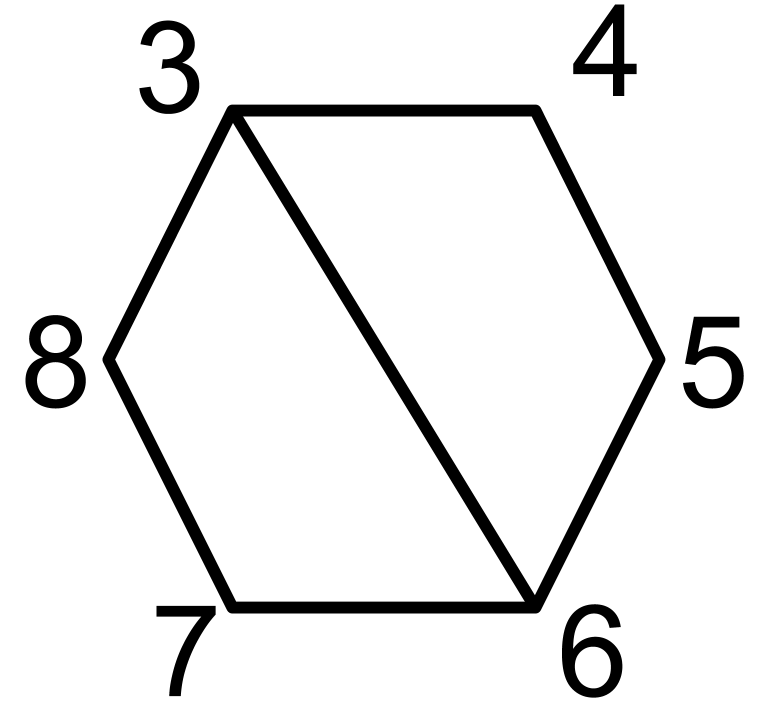} \nonumber \\
   =  \left( \frac{1}{X_{1,4}} + \frac{1}{X_{3,6}} \right)
   &\left( \frac{1}{X_{1,6}} + \frac{1}{X_{3,8} - X_{3,6}} \right) +
   \left( \frac{1}{X_{1,4}} + \frac{1}{X_{3,8}} \right)
   \left( \frac{1}{X_{5,8}} + \frac{1}{X_{3,6} - X_{3,8}} \right)
\end{align}

We can illustrate the application of \eref{eq:stokesrecursion} for $n = 10$ on the following example, 

\begin{align}
    \underline{\Omega} \addpic{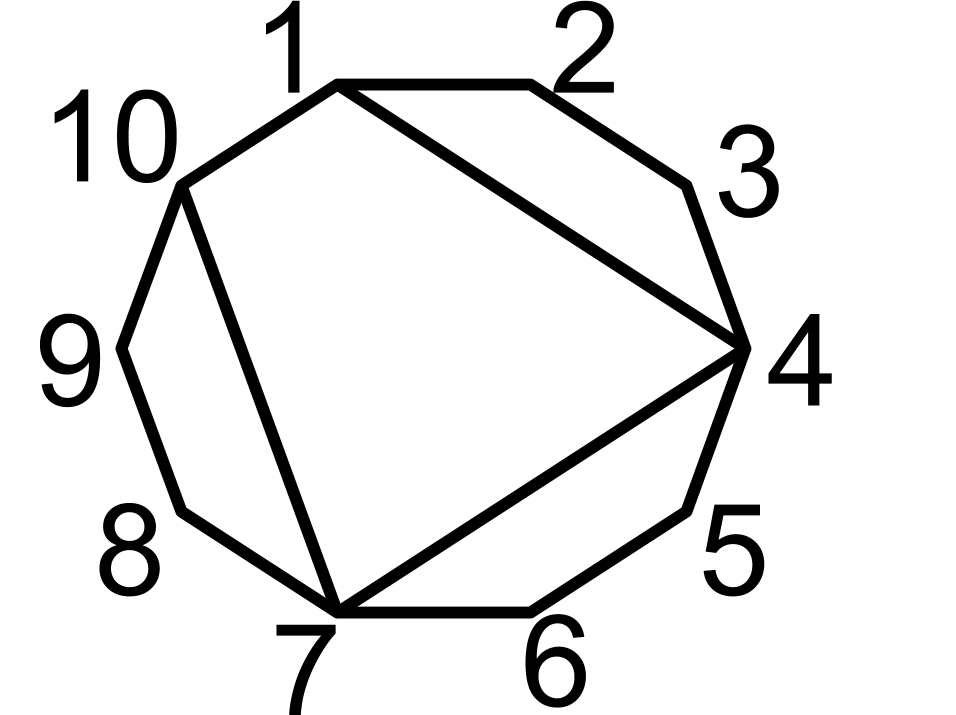} = \left( \frac{1}{X_{1,4}} + \frac{1}{X_{3,6}} \right) D_{X_{3,6}} \underline{\Omega} \addpic{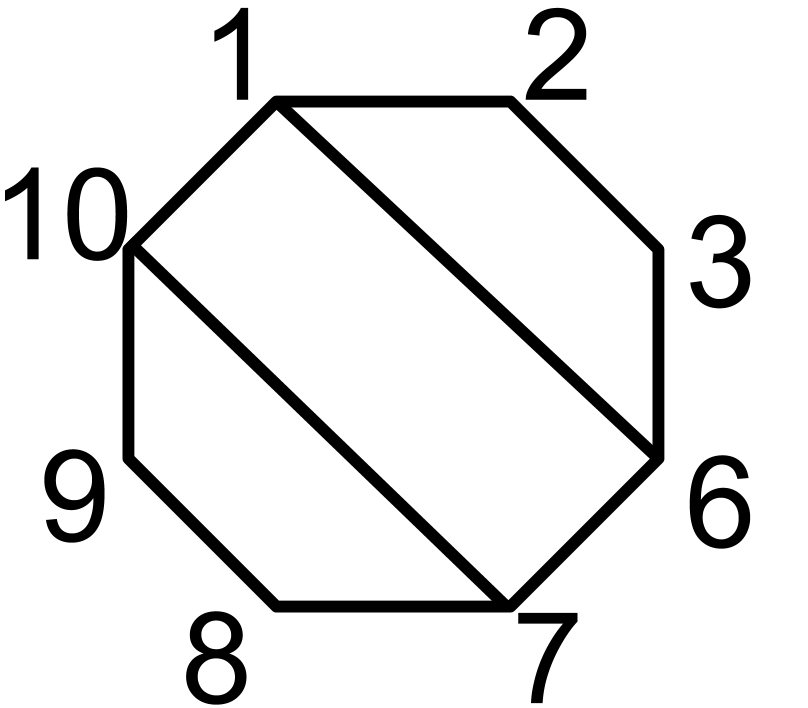}
    D_{X_{3,6}} \underline{\Omega} \addpic{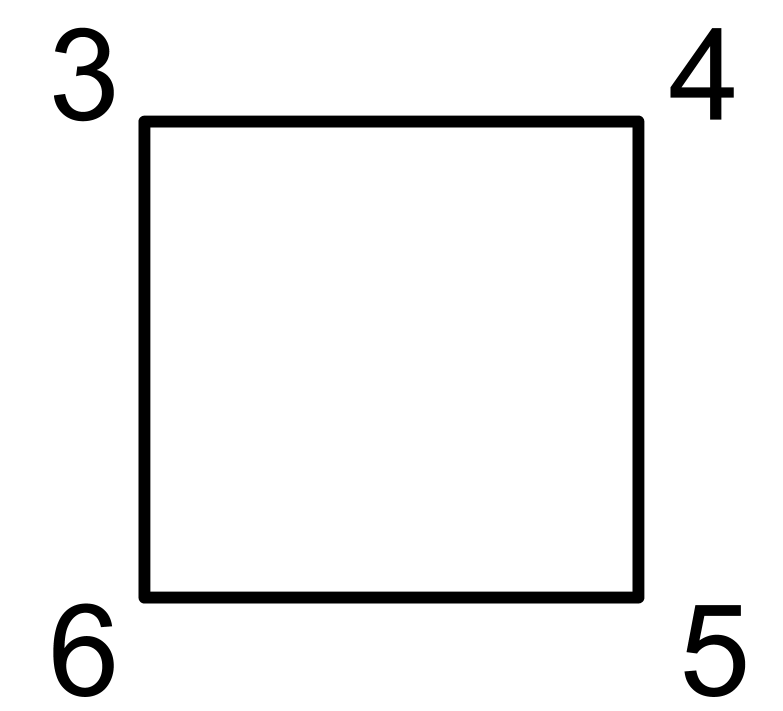} + \nonumber \\
    + \left( \frac{1}{X_{1,4}} + \frac{1}{X_{3,10}} \right) D_{X_{3,10}}
     \underline{\Omega} \addpic{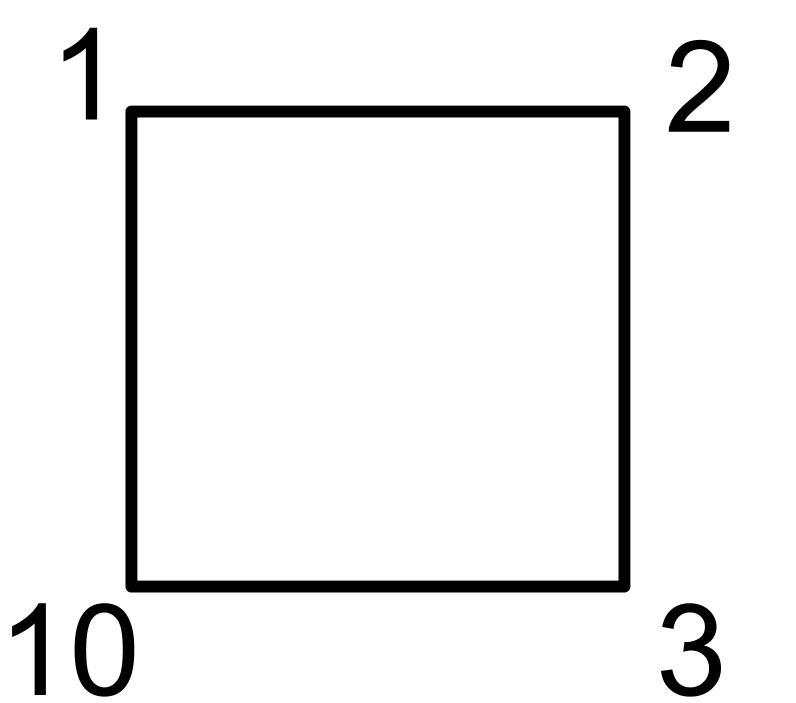} D_{X_{3,10}} \underline{\Omega} \addpic{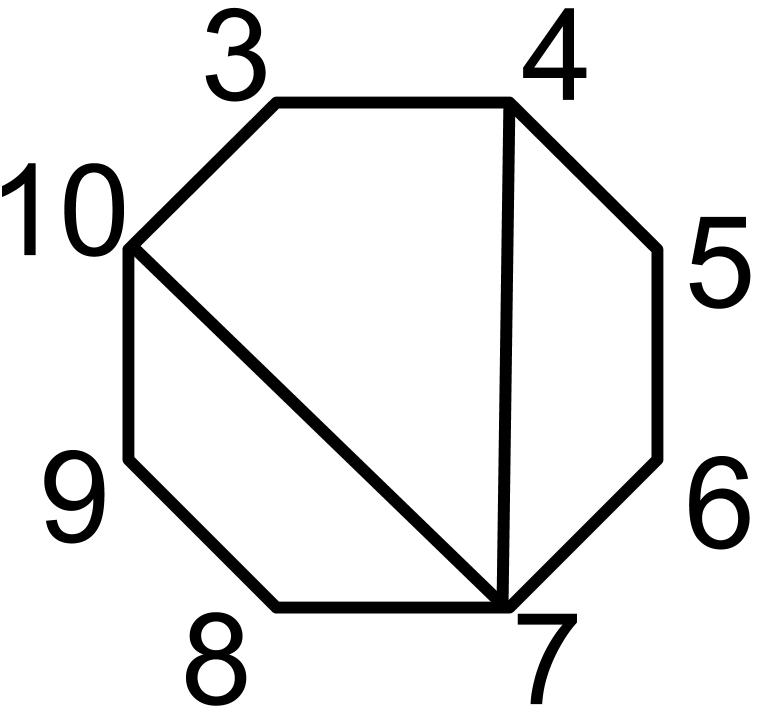}
\end{align}

\subsection{Assembling the amplitude}

As mentioned in the beginning of this section, in order to obtain the full amplitude in $\phi^4$ theory one has to sum the contributions coming from \emph{all} Stokes polytopes. Furthermore, each contribution has to be weighted by an appropriate factor which takes into account the over counting of Feynman diagrams shared by multiple Stokes polytope. While correct, this is not a completely satisfactory result because ultimately requires to sum over all Feynman diagrams (one per Stokes polytope). 
On the other hand, our recursive formulae suggest a simpler way to avoid overcounting of Feynman diagrams. First note that taking the limit $X_{i,j}\to 0$ in the canonical form of a Stokes polytope $S_Q$ kills all the contributions coming from Feynman diagrams with a propagator $X_{i,j}$. Looking back at \eqref{eq:stokesrecursion} two particularly simple choices are the limits $X_{1,4} \to 0$ or $X_{3,i} \to 0$, it is then natural to ask whether it is possible to obtain the full amplitude by summing the contributions coming from \emph{some} Stokes polytopes, on each of which a suitable limit is taken. While we have no proof of the general case, we offer some low multiplicity examples where this is possible.

As a warmup exercise, let us consider the case of $n = 6$. For orientations of diagonals defined by assigning '+' to odd and '-' to even vertices, it is easy to check that diagonals $X_{1,4}$ and $X_{3,6}$ are compatible with $X_{1,4}$; $X_{2,5}$ and $X_{1,4}$ are compatible with $X_{2,5}$.
Let us then write the amplitude in the form

\begin{align}
    A_6 = \frac{1}{X_{1,4}} + \frac{1}{X_{2,5}} +\frac{1}{X_{3,6}} = \underline{\Omega} \addpic{Hexagon14.png}  + \lim_{X_{1,4} \rightarrow \infty} \underline{\Omega} \addpic{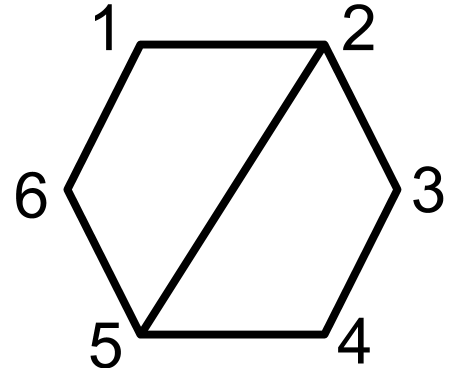}
\end{align}

 In order to piece together the amplitude for $n = 8$, in addition to \eref{eq:8ptbox}, \eref{eq:8ptpentagon} we will need to consider three more Stokes polytopes. Those functions are obtained by picking $X_b = X_{1,4}$, $X_b = X_{3,8}$, $X_b = X_{1,6}$, respectively, and are given by

\begin{align}
\label{eq:pentabox2}
    &\underline{\Omega}\addpic{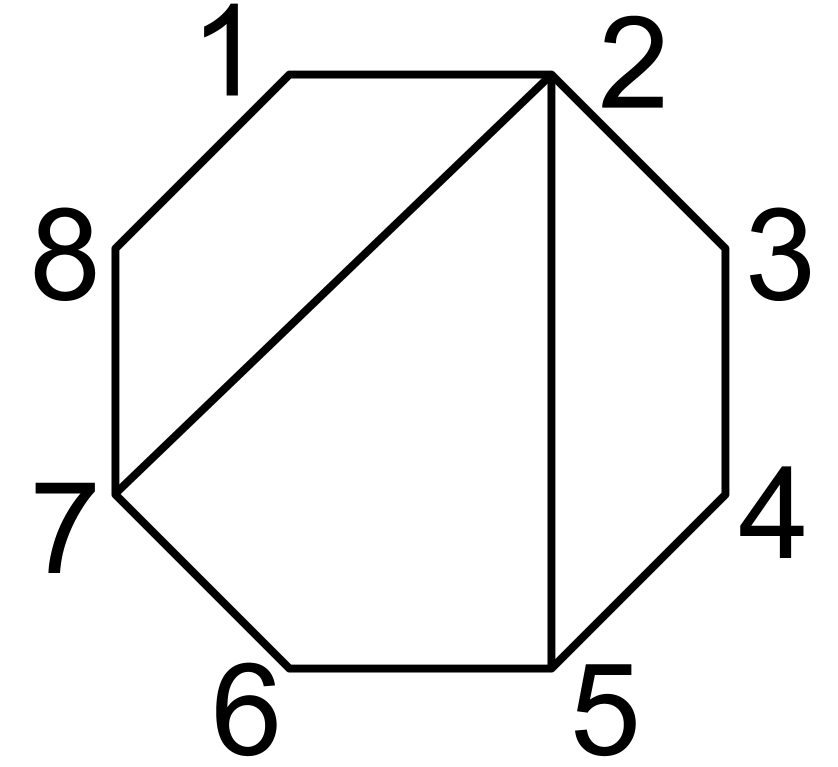} = \left( \frac{1}{X_{1,4}} + \frac{1}{X_{2,5}} \right)
   \left( \frac{1}{X_{1,6}} + \frac{1}{X_{2,7} - X_{2,5}} \right) +
   \left( \frac{1}{X_{1,4}} + \frac{1}{X_{2,7}} \right)
   \left( \frac{1}{X_{4,7}} + \frac{1}{X_{2,5} - X_{2,7}} \right) \nonumber \\
   &\underline{\Omega}\addpic{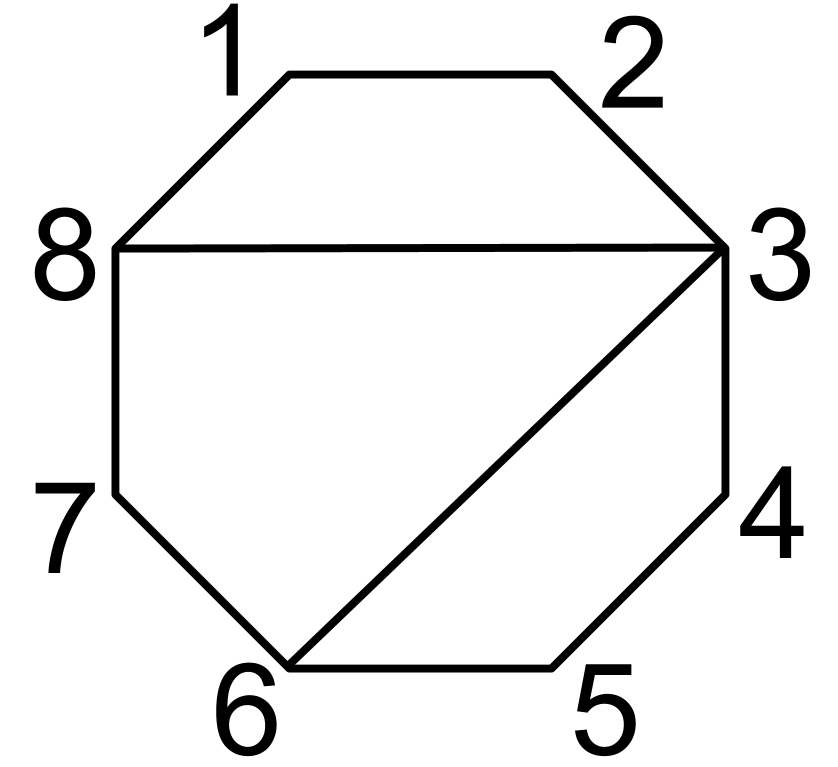} = \left( \frac{1}{X_{3,8}} + \frac{1}{X_{2,5}} \right)
   \left( \frac{1}{X_{5,8}} + \frac{1}{X_{2,7} - X_{2,5}} \right) +
   \left( \frac{1}{X_{3,8}} + \frac{1}{X_{2,7}} \right)
   \left( \frac{1}{X_{3,6}} + \frac{1}{X_{2,5} - X_{2,7}} \right) \nonumber \\
   &\underline{\Omega}\addpic{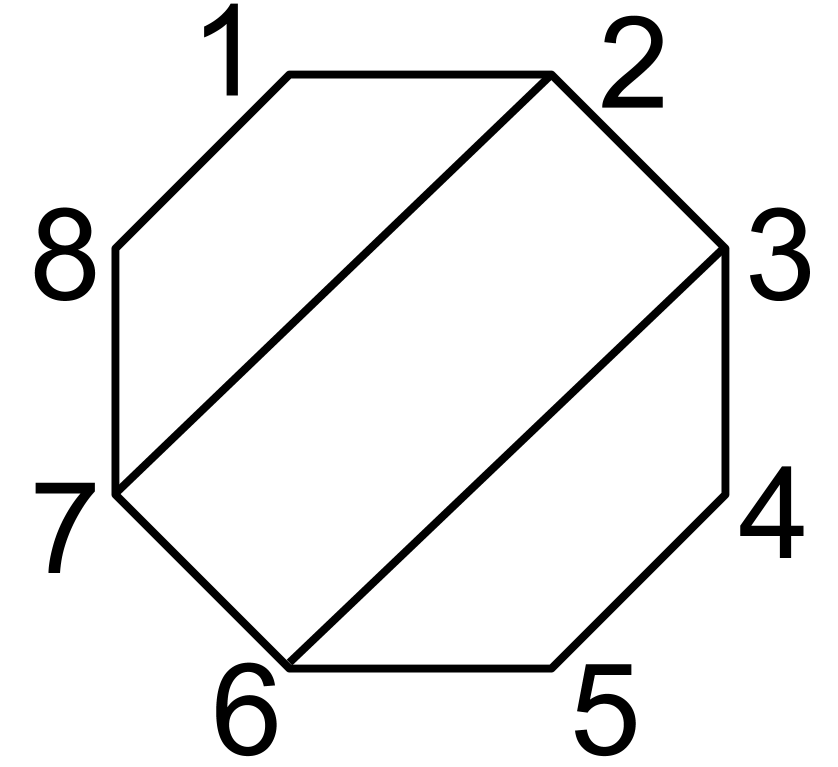} =  \left( \frac{1}{X_{1,6}} + \frac{1}{X_{2,7}} \right) 
   \left( \frac{1}{X_{2,5}} + \frac{1}{X_{3,6}}  \right)
\end{align}

 Then, the 8-particle amplitude can be obtained as

\begin{align}
\label{eq:8pamp}
    A_8 = \underline{\Omega}\addpic{PentagonType.png} + \lim_{X_{5,8} \rightarrow \infty} \underline{\Omega}\addpic{BoxType.png} + \lim_{X_{1,4} \rightarrow \infty} \underline{\Omega}\addpic{PentagonType1.png} + \nonumber \\
    + \lim_{X_{3,8} \rightarrow \infty} \underline{\Omega}\addpic{PentagonType2.png} - \lim_{X_{3,6} \rightarrow \infty} \lim_{X_{1,6 \rightarrow \infty}} \underline{\Omega} \addpic{BoxType2.png}
\end{align}

A nice feature of this formula is that the limits taken in Eq.~(\ref{eq:8pamp}) match the $X_b$ facets of Eq.~(\ref{eq:8ptbox}), Eq.~(\ref{eq:8ptpentagon}), Eq.~(\ref{eq:pentabox2}). The fact that the facet $X_b$ appears only in the prefactors of the recursive formula Eq.~(\ref{eq:stokesrecursion}) results in a simple expression for the amplitude.

\section{Conclusions}
\label{sec:conclusions}

In this paper we presented a simple way to compute recursively both the canonical form and the canonical rational function associated to simple polytopes, motivated by the desire of making manifest the projectivity of the former. As it is familiar from the case of BCFW formulae \cite{Britto:2005fq}, this is done at the cost of introducing spurious poles.

Compared to more standard recursive formulae, such as those in \cite{He:2018svj}, based on triangulations which do not involve other vertices other than those of the polytope itself, our formula have the advantage of introducing only linear spurious poles\footnote{We would like to thank Song He for pointing at us the importance of this fact.}. This fact might turn out to be helpful especially in the case of polytopes associated to integrands when addressing the loop integration.
Furthermore, it is not always obvious how to uplift identities which are true on the subspace where a given polytope is realized to the space of all its facet variables. Our formulae, however, are proved to hold directly in this space. In particular, although it is important to know that a convex realization of the polytope exists, one can setup recursive formulae without the explicit knowledge of such realization.

For illustrational purposes we applied our results to Stokes polytopes, which are known to be related to a planar theory with quartic interactions\footnote{By applying our results to the so called ABHY Associahedra - the polytopes associated to tree and 1-loop level amplitudes in $\phi^3$ bi-adjoint theory - one recovers as a special case certain recursive formulae which were originally derived using a remarkable projection property satisfied by these polytopes \cite{GiulioNima, GiulioPhD}.}.
A convex realization of these polytopes was presented in \cite{Manneville:2018}, see also \cite{Aneesh:2019cvt} for connections with ABHY associahedra. As already discussed, however, we can write down recursive formulae for both the canonical forms and the rational functions of Stokes polytope without making explicit use of these geometric data. Also, the structure of the recursion makes manifest how to perform certain limits on the rational function of Stokes polytopes which could be used to assemble them together in a planar quartic amplitudes in a more efficient way.

An interesting direction for further investigation would be to consider canonical forms of non simple polytopes. 
Dually one has to consider volumes of non simplicial polytopes which can be computed by iterating the standard triangulation method described in \cite{Arkani-Hamed:2017mur} for all non simplicial facets. 
\begin{figure}[h!]
    \centering
    \includegraphics[width=0.45\textwidth]{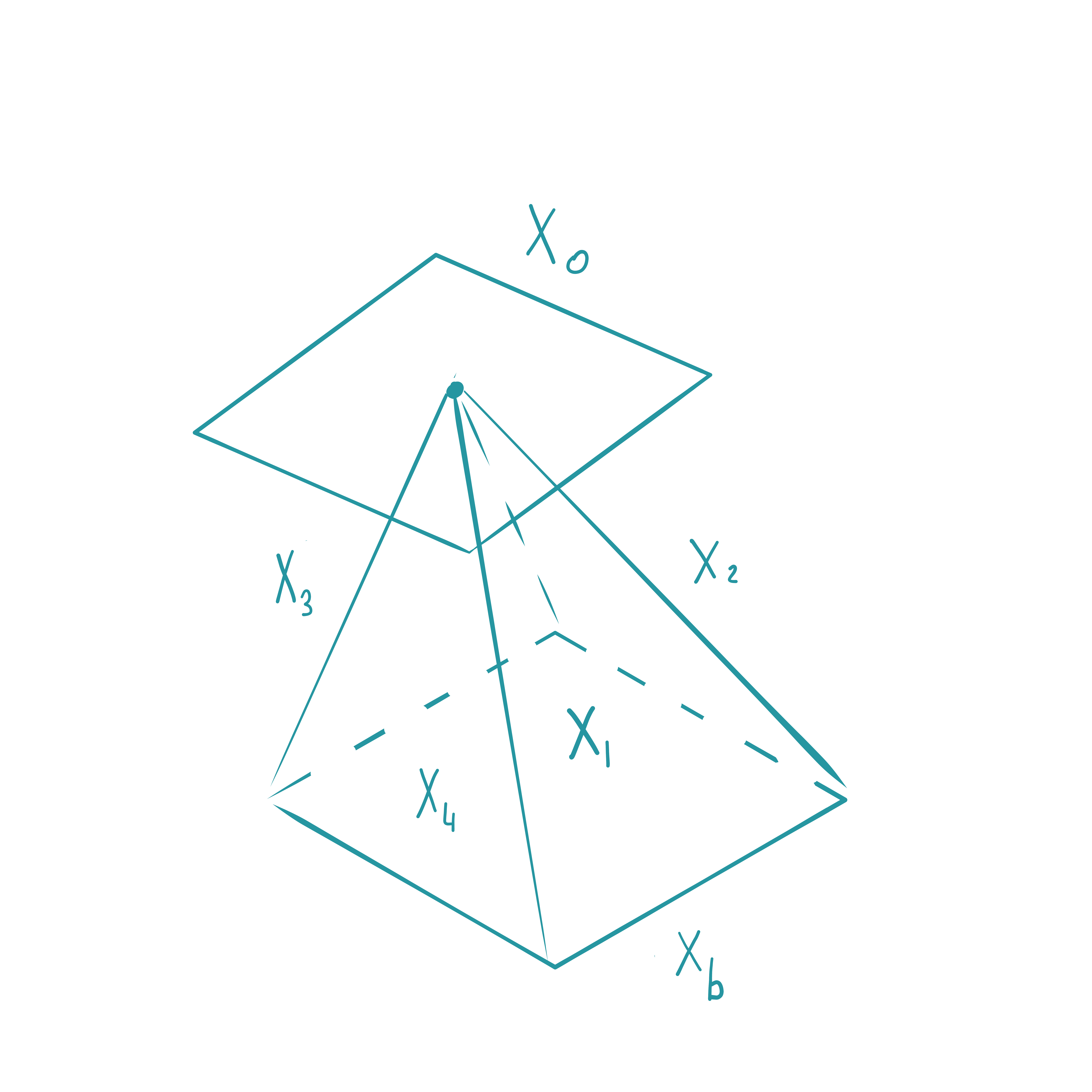}
    \caption{A square pyramid is not simple because the four facets $X_1,X_2,X_3$ and $X_4$ meet at a common vertex. The standard triangulation of the dual involves the external facet $X_0$ in addition to the plane at infinity.}
    \label{fig:Pyramid}
\end{figure}

The simplest example is that of the pyramid shown in Fig. \ref{fig:Pyramid}, whose canonical form is given by
\begin{align}
    \Omega = \dlog\left(\frac{X_b}{X_0}\right) \wedge \dlog\left(\frac{X_1}{X_3}\right) \wedge \left.\dlog\left(\frac{X_2}{X_4} \right)\right|_H,
\end{align}
where $H$ is the subspace defined by $\{ X_2 =X_1+ \alpha X_3 + \beta X_0,\      X_4 = X_1+ \alpha' X_3 + \beta' X_0\}$.
The novelty with respect to \ref{eq:canonicalsimple2} is that we cannot think of facet $\Omega$ as living in the space of independent variables otherwise $\Omega$ would develop a spurious pole along the plane $X_0$. Because of this a generalization of our formula is not immediate and we leave it for a future work.

\section{Acknowledgments}

The authors would like to thank Frederic Chapoton for help in understanding the combinatorical structure of Stokes polytopes. GS would like to thank Nima Arkani-Hamed, Song He and Hadleigh Frost for stimulating discussions on several topics around polytopes as well as R. Obrist for support during a preliminary part of this work. SS would like to thank Marcus Spradlin for stimulating discussions and support. GS is supported by the Simons Investigator Award\#376208 of A. Volovich. SS was supported in part by DOE grant DE-SC0010010 Task A.

\appendix

\section{Explicit derivation of the recursive formula for $\Omega$}\label{app:explicit}

In this Appendix we provide an alternative, and somewhat more explicit, proof of our recursive formula. 
We focus on each of the vertices of our polytope $\mathcal{P}$ and show that every term in \eref{eq:canonicalsimple} is reproduced by a sum of the corresponding terms in \eref{eq:recursionform}. Like in the derivation of Section \ref{sec:omegarec}, there will be several types of vertices to consider.

\begin{enumerate}
    \item Vertex $v \notin X_b$, such that $(v,B) = \bigcap_{a=1}^{d-1} X_{s_a}$, for some vertex $B \in X_b$ and a collection of $d-1$ facets $X_{s_a} \in \mathcal{S}$. Here, $v$ will belong to only one element of $\mathcal{U}$, which we may denote as $X_{u_i}$. Let the corresponding terms of $\Omega$ evaluated as in \eref{eq:canonicalsimple} be
    
    \begin{align}
    \label{eq:a1}
         \Omega_v = d \log X_{u_i} \bigwedge_{a=1}^{d-1} d \log X_{s_a} \nonumber \\
         \Omega_B = - d \log X_b \bigwedge_{a=1}^{d-1} d \log X_{s_a}
    \end{align}
    
    Here, we have chosen the ordering of $X_{s_a}$ in the wedge product in such a way as to absorb any possible overall minus sign in the first line. There is only one term in \eref{eq:recursionform} to consider, giving
    
    \begin{align}
        \tilde{\Omega}_v = d \log X_{u_i} \bigwedge_{a=1}^{d-1} d \log X_{s_a} -  d \log X_b \bigwedge_{a=1}^{d-1} d \log X_{s_a}
    \end{align}
    
    producing precisely the sum of the two terms of \eref{eq:a1}. Note that the relative minus sign between two terms is in accordance with the sign flip rule as described in the Section \ref{sec:polytopes}.
    
    \item Vertex $v \notin X_b$, such that $(v,U) = \bigcap_{a=1}^{d-1} X_{s_a}$, $X_{s_a} \in \mathcal{S}$, where $v$ belongs to some element $X_{u_i}$ of $\mathcal{U}$ and $U$ belongs to some element $X_{u_j}$ of $\mathcal{U}$. When we were studying the recursive formula for canonical rational functions, this is precisely the choice of $\mathcal{U}$ we aimed to avoid. When working with canonical forms, however, signs conspire in such a way that the troublesome term is canceled between $v$ and $U$. Let the sum of terms of \eref{eq:canonicalsimple} corresponding to $v$ and $U$ (written in accordance with the sign flip rule) be
    
    \begin{align}
    \label{eq:vtermsimple}
        \Omega_v + \Omega_U = d \log X_{u_i} \bigwedge_{a=1}^{d-1} d \log X_{s_a} - d \log X_{u_j} \bigwedge_{a=1}^{d-1} d \log X_{s_a}
    \end{align}
    
    Note that we have again chosen the ordering of $X_{s_a}$ in the wedge product in such a way that any potential minus sign of the first term is absorbed. This expression is to be compared with the sum of terms corresponding to vertices $v$ and $U$ on the right hand side of \eref{eq:recursionform},
    
    \begin{align}
    \label{eq:vtermrec}
        \tilde{\Omega}_v + \tilde{\Omega}_U = \left( d \log X_{u_i} \bigwedge_{a=1}^{d-1} d \log X_{s_a} - d \log X_b \bigwedge_{a=1}^{d-1} d \log X_{s_a} \right) - \nonumber \\ \left( d \log X_{u_j} \bigwedge_{a=1}^{d-1} d \log X_{s_a} - d \log X_b \bigwedge_{a=1}^{d-1} d \log X_{s_a} \right)
    \end{align}
    
    The relative sign of the two terms has to be a minus sign, in order to ensure the projectivity of form $\Omega$ on the line $\bigcap_{a=1}^{d-1} X_{s_{a}}$. To say it differently, the minus sign comes from the flip rule described in Section \ref{sec:polytopes}, considering that a mutation of $X_{u_i}$ takes us from vertex $v$ to vertex $U$. Now the unwanted $X_b$ - dependent term cancels and we can see that $\Omega_v + \Omega_U = \tilde{\Omega}_v + \tilde{\Omega}_U$.
    
    \item Vertex $v \notin X_b$, such that $v$ belongs to some collection of facets in $\mathcal{U}$, $X_{u_1} , \hdots , X_{u_n}$. Let the corresponding term in \eref{eq:canonicalsimple} be
    
    \begin{align}
        \Omega_v = \bigwedge_{i=1}^n d \log X_{u_i} \bigwedge_{a=1}^{d-n} d \log X_{s_a} 
    \end{align}
    
    Then, we will need to consider the sum of $n$ terms in \eref{eq:recursionform}, giving us
    
    \begin{align}
    \tilde{\Omega}_v = 
        \sum_{i=1}^{n} \bigwedge_{j = 1}^{i-1} d \log \left( X_{u_j} - X_{u_i} \right) \wedge d \log X_{u_i} \bigwedge_{k = i + 1}^n d \log \left( X_{u_k} - X_{u_i} \right) \bigwedge_{a=1}^{d-n} d \log X_{s_a} - \nonumber \\
        \sum_{i=1}^{n} \bigwedge_{j = 1}^{i-1} d \log \left( X_{u_j} - X_{u_i} \right) \wedge d \log X_{b} \bigwedge_{k = i + 1}^n d \log \left( X_{u_k} - X_{u_i} \right) \bigwedge_{a=1}^{d-n} d \log X_{s_a}
    \end{align}
    
   Now, we can see that the second line vanishes after expanding out the wedge products and using the identity in \eref{eq:partialfrac2}. Similarly, using the identity in \eref{eq:partialfrac}, we can see that the first line simplifies to the form $\tilde{\Omega}_v = \Omega_v$.

    \item Vertex $v \in X_b$, such that $v$ belongs to only one element of $\mathcal{U}$, which we will denote as $X_{u_i}$. Let us write the corresponding term of $\Omega$ as
    
    \begin{align}
        \Omega_v = d \log X_{u_i} \wedge d \log X_{b} \bigwedge_{j=1}^{n-2} d \log X_{s_j}
    \end{align}
    
    Then, there is only one term in \eref{eq:recursionform} to consider, giving us
    
    \begin{align}
        \tilde{\Omega}_v = (d \log X_{u_i} - d \log X_b) \wedge d \log (X_b + X_{u_i}) \bigwedge_{j=1}^{n-2} d \log X_{s_j} = \Omega_v
    \end{align}
    
    where we used the identity
    
    \begin{align}
    \label{eq:ourfavorite}
        d \log \left( \frac{a}{b} \right) \wedge d \log(a + b) = d \log a \wedge d \log b
    \end{align}
    
    \item Vertex $v \in X_b$, such that $v$ belongs to some collection of $n$ elements of $\mathcal{U}$, denoted as $X_{u_1}$, $X_{u_2}$, $\hdots$, $X_{u_n}$. Let us write the corresponding term of $\Omega$ as
    
    \begin{align}
        \Omega_v = d \log X_b \bigwedge_{i=1}^n d \log X_{u_i} \bigwedge_{a=1}^{d-n-1} d \log X_{s_a}
    \end{align}
    
    Now, there are $n$ terms in \eref{eq:recursionform} to consider, which after using the identity in  \eref{eq:ourfavorite} give 
    
    \begin{align}
       \tilde{\Omega}_v =  \sum_{i=1}^{n} d \log X_b \bigwedge_{j = 1}^{i-1} d \log \left( X_{u_j} - X_{u_i} \right) \wedge d \log X_{u_i} \bigwedge_{k = i + 1}^n d \log \left( X_{u_k} - X_{u_i} \right) \bigwedge_{a=1}^{d-n-1} d \log X_{s_a}
    \end{align}
    
    which after expanding the wedge products and using the identity in \eref{eq:partialfrac2}, in the same manner as before, gives $\tilde{\Omega}_v = \Omega_v$.
    
\end{enumerate}


\begin{thebibliography}{}

%\cite{Arkani-Hamed:2013jha}
\bibitem{Arkani-Hamed:2013jha} 
  N.~Arkani-Hamed and J.~Trnka,
  %``The Amplituhedron,''
  JHEP {\bf 1410}, 030 (2014)
  doi:10.1007/JHEP10(2014)030
  [arXiv:1312.2007 [hep-th]].
  %%CITATION = doi:10.1007/JHEP10(2014)030;%%
  %243 citations counted in INSPIRE as of 21 Nov 2019
  
  %\cite{ArkaniHamed:2012nw}
\bibitem{ArkaniHamed:2012nw} 
  N.~Arkani-Hamed, J.~L.~Bourjaily, F.~Cachazo, A.~B.~Goncharov, A.~Postnikov and J.~Trnka,
  %``Grassmannian Geometry of Scattering Amplitudes,''
  doi:10.1017/CBO9781316091548
  arXiv:1212.5605 [hep-th].
  %%CITATION = doi:10.1017/CBO9781316091548;%%
  %307 citations counted in INSPIRE as of 10 Dec 2019
  
  %\cite{Arkani-Hamed:2017mur}
\bibitem{Arkani-Hamed:2017mur} 
  N.~Arkani-Hamed, Y.~Bai, S.~He and G.~Yan,
  %``Scattering Forms and the Positive Geometry of Kinematics, Color and the Worldsheet,''
  JHEP {\bf 1805}, 096 (2018)
  doi:10.1007/JHEP05(2018)096
  [arXiv:1711.09102 [hep-th]].
  %%CITATION = doi:10.1007/JHEP05(2018)096;%%
  %54 citations counted in INSPIRE as of 21 Nov 2019
  
  %\cite{Britto:2005fq}
\bibitem{Britto:2005fq} 
  R.~Britto, F.~Cachazo, B.~Feng and E.~Witten,
  %``Direct proof of tree-level recursion relation in Yang-Mills theory,''
  Phys.\ Rev.\ Lett.\  {\bf 94}, 181602 (2005)
  doi:10.1103/PhysRevLett.94.181602
  [hep-th/0501052].
  %%CITATION = doi:10.1103/PhysRevLett.94.181602;%%
  %1053 citations counted in INSPIRE as of 06 Dec 2019
  
  %\cite{Banerjee:2018tun}
\bibitem{Banerjee:2018tun} 
  P.~Banerjee, A.~Laddha and P.~Raman,
  %``Stokes polytopes: the positive geometry for $\phi^{4}$ interactions,''
  JHEP {\bf 1908}, 067 (2019)
  doi:10.1007/JHEP08(2019)067
  [arXiv:1811.05904 [hep-th]].
  %%CITATION = doi:10.1007/JHEP08(2019)067;%%
  %7 citations counted in INSPIRE as of 21 Nov 2019

%\cite{Manneville:2018}
\bibitem{Manneville:2018}  
 T.~Manneville and V.~Pilaud, Discrete \& Computational Geometry 61, 507 (2018)
 doi:10.1007/s00454-018-0004-2
 [arXiv:1703.09953 [math.CO]].
 
 %\cite{Arkani-Hamed:2017tmz} 
\bibitem{Arkani-Hamed:2017tmz} 
  N.~Arkani-Hamed, Y.~Bai and T.~Lam,
  %``Positive Geometries and Canonical Forms,''
  JHEP {\bf 1711}, 039 (2017)
  doi:10.1007/JHEP11(2017)039
  [arXiv:1703.04541 [hep-th]].
  %%CITATION = doi:10.1007/JHEP11(2017)039;%%
  %48 citations counted in INSPIRE as of 02 Dec 2019
  
  %\cite{Aneesh:2019cvt}
\bibitem{Aneesh:2019cvt} 
  P.~B.~Aneesh, P.~Banerjee, M.~Jagadale, R.~Rajan, A.~Laddha and S.~Mahato,
  %``On Positive Geometries of Quartic Interactions II : Stokes polytopes, Lower Forms on Associahedra and Worldsheet Forms,''
  arXiv:1911.06008 [hep-th].
  %%CITATION = ARXIV:1911.06008;%%
  %1 citations counted in INSPIRE as of 02 Dec 2019
 
%\cite{Baryshnikov}  
\bibitem{Baryshnikov}
  Y.~Baryshnikov, New Developments in Singularity Theory 65 (2001)
  % On Stokes sets
  doi:10.1007/978-94-010-0834-$1_3$.
  
%\cite{Chapoton}  
\bibitem{Chapoton}
  F.~Chapoton,
  %Stokes posets and serpent nests
  Discrete Mathematics \& Theoretical Computer Science 18(3) (2015) 
  arXiv:1505.05990 [math.RT]
  
%\cite{PaluPilaud}
\bibitem{PaluPilaud}
Y.~Palu, V.~Pilaud and P.~Plamondon, arXiv:1707.07574 [math.CO].
  
%\cite{He:2018svj}
\bibitem{He:2018svj} 
  S.~He and Q.~Yang,
  %``An Etude on Recursion Relations and Triangulations,''
  JHEP {\bf 1905}, 040 (2019)
  doi:10.1007/JHEP05(2019)040
  [arXiv:1810.08508 [hep-th]].
  %%CITATION = doi:10.1007/JHEP05(2019)040;%%
  %6 citations counted in INSPIRE as of 05 Dec 2019
  
  %\cite{Salvatori:2018aha}
\bibitem{Salvatori:2018aha} 
  G.~Salvatori,
  %``1-loop Amplitudes from the Halohedron,''
  arXiv:1806.01842 [hep-th].
  %%CITATION = ARXIV:1806.01842;%%
  %18 citations counted in INSPIRE as of 06 Dec 2019
  
  %\cite{Raman:2019utu}
\bibitem{Raman:2019utu} 
  P.~Raman,
  %``The positive geometry for $\phi^{p}$ interactions,''
  JHEP {\bf 1910}, 271 (2019)
  doi:10.1007/JHEP10(2019)271
  [arXiv:1906.02985 [hep-th]].
  %%CITATION = doi:10.1007/JHEP10(2019)271;%%
  %6 citations counted in INSPIRE as of 06 Dec 2019
  
  %\cite{Jagadale:2019byr}
\bibitem{Jagadale:2019byr} 
  M.~Jagadale, N.~Kalyanapuram and A.~Prema Balakrishnan,
  %``Accordiohedra as Positive Geometries for Generic Scalar Field Theories,''
  arXiv:1906.12148 [hep-th].
  %%CITATION = ARXIV:1906.12148;%%
  %3 citations counted in INSPIRE as of 06 Dec 2019
  
  %\cite{Gao:2017dek}
\bibitem{Gao:2017dek} 
  X.~Gao, S.~He and Y.~Zhang,
  %``Labelled tree graphs, Feynman diagrams and disk integrals,''
  JHEP {\bf 1711}, 144 (2017)
  doi:10.1007/JHEP11(2017)144
  [arXiv:1708.08701 [hep-th]].
  %%CITATION = doi:10.1007/JHEP11(2017)144;%%
  %24 citations counted in INSPIRE as of 06 Dec 2019

\bibitem{GiulioNima}
N.~Arkani-Hamed, S.~He, G.~Salvatori and H.~Thomas, [arXiv:1912.12948 [hep-th]].

\bibitem{GiulioPhD}
G. Salvatori, [http://hdl.handle.net/2434/740134]

  
 \bibitem{orlik}
 P. Orlik and H. Terao, Arrangements of hyperplanes, Springer.
 
\end{thebibliography}
\end{document}